\documentclass[11pt]{article}
\usepackage{graphicx}
\usepackage{subcaption} 
\usepackage{caption}
\usepackage{float} 
\usepackage{a4wide}
\usepackage{mathrsfs}
\usepackage{latexsym,bm}
\usepackage{graphicx}
\usepackage{indentfirst}
\usepackage{slashed}
\usepackage{amsmath}
\usepackage{amssymb}
\usepackage[usenames,dvipsnames]{color}
\usepackage{hyperref}
\usepackage{epsfig}
\usepackage[titletoc]{appendix}
\usepackage{multirow}%
\usepackage{rotating}
\usepackage[square,numbers,sort&compress]{natbib}
\usepackage{placeins}   
\usepackage{ulem} 
\usepackage[top=2.9cm,bottom=2.5cm,left=2.8cm,right=3cm]{geometry}
\usepackage{tabularx}

\usepackage{booktabs}  
\usepackage{siunitx}   
\usepackage{amsmath}  

\renewcommand{\arraystretch}{1.0}

\newcommand{\email}[1]{\footnote{{\em } \texttt{#1}}}

\newcommand{\bma}{\left(\begin{matrix}}
\newcommand{\ema}{\end{matrix}\right)}

\newcommand{\xpt}{{\chi}{\rm PT}}

\title{\Large \bf Doubly charmed baryon-light meson scattering in chiral effective theory with lattice constraints }
\author{\small Peng-Qi Wang$^a$,\, Zhi-Hui Guo$^{a}$\email{zhguo@hebtu.edu.cn} \\[0.5em]
{ \small\it ${}^a$ Department of Physics and Hebei Key Laboratory of Photophysics Research and Application, } \\ 
{\small\it Hebei Normal University,  Shijiazhuang 050024, China}
}
\date{}

\begin{document}

\maketitle
\begin{abstract}
We study the scattering of the ground states of doubly charmed baryons ($\Xi_{cc}^{++},\Xi_{cc}^{+},\Omega_{cc}^{+}$) and light-flavor pseudoscalar mesons ($\pi,K,\eta$) up to the next-to-leading order within chiral effective theory. We perform the unitarization of the $S$-wave scattering amplitudes in order to study the excited doubly charmed baryons. The unknown next-to-leading order low energy constants are determined through the fits to recent lattice data in the elastic scattering processes based on the CLQCD ensembles. Following the chiral extrapolation to physical quark masses, we predict resonance, virtual and bound doubly-charmed-baryon states arising from the single- and coupled-channel scattering of $\Xi_{cc}^{++},\Xi_{cc}^{+},\Omega_{cc}^{+}$ with $\pi,K,\eta$. Furthermore, we also calculate the corresponding scattering lengths, effective ranges, phase shifts and inelasticities at physical quark masses, which could shed light on future experimental searches and lattice simulations.  
\end{abstract}

\section{Introduction}

Doubly charmed baryons (DCBs), composed of two charm quarks and one light quark, are important objects for investigating the nonperturbative dynamics of the strong interaction~\cite{Korner:1994nh,Kiselev:2001fw,Ma:2003zk,Hu:2005gf,Chang:2006xp,Chang:2006eu,Ebert:2002ig,Karliner:2014gca,Lewis:2001iz,Liu:2009jc,Brown:2014ena}. Compared to singly charmed baryons, DCBs contain two heavy quarks and one light quark simultaneously. The presence of two heavy quarks brings the system closer to the heavy quark limit~\cite{Hu:2005gf}, further enhancing the dominance of heavy quark degrees of freedom in low-energy dynamics. This makes them important objects for testing heavy quark symmetry and chiral dynamics.

In recent years, with the successive development of high-energy physics experiments, significant progresses have been achieved for the research on DCBs. In 2017, the LHCb collaboration observed the DCB $\Xi^{++}_{cc}$ through the decay $\Xi^{++}_{cc}\rightarrow\Lambda^{+}_{c}K^{-}\pi^+\pi^+$, measuring its mass to be $3621.40\pm0.78\,\mathrm{MeV}$~\cite{LHCb:2017iph}. Then $\Xi^{++}_{cc}$ was measured in another independent process $\Xi^{++}_{cc}\rightarrow\Xi^{+}_{c}\pi^+$ by LHCb, with the mass of $3620.6\pm1.5 \,\mathrm{MeV}$~\cite{LHCb:2018pcs}, which is in excellent agreement with its previous result in Ref.~\cite{LHCb:2017iph}. Following these discoveries, LHCb has performed high-precision measurements of the lifetime~\cite{LHCb:2018zpl}, mass~\cite{LHCb:2019epo}, and production cross-section~\cite{LHCb:2019qed} of $\Xi^{++}_{cc}$, and also observed a new decay mode $\Xi^{++}_{cc}\rightarrow\Xi^{0}_{c}\pi^+\pi^+$~\cite{LHCb:2025shu}, providing reliable benchmarks for theoretical studies. Recently, the LHCb collaboration has reported the observation of $\Xi^{+}_{cc}$ via the decay $\Xi^{+}_{cc}\rightarrow\Lambda^{+}_{c}K^{-}\pi^{+}$, with a mass of $3619.97\pm0.83\pm0.26^{+1.90}_{-1.30}\,\mathrm{MeV}$ and a statistical significance exceeding $7\sigma$~\cite{LHCb:2026pxn}. This measurement confirms the $\Xi^{+}_{cc}$ as the isospin partner of the $\Xi^{++}_{cc}$ and completes the isospin doublet of doubly charmed baryons. The mass of $\Xi^{+}_{cc}$ measured by LHCb is compatible with many theoretical predictions~\cite{Ebert:2002ig,Karliner:2014gca,Korner:1994nh,Lewis:2001iz,Liu:2009jc,Brown:2014ena}, in contrast to the previous result from 
the SELEX collaboration~\cite{SELEX:2002wqn,SELEX:2004lln} that showed significant discrepancies with both the LHCb measurement and the theoretical expectations. 
The valence quark compositions of $\Xi^{++}_{cc}$ and $\Xi^{+}_{cc}$ are $ccu$ and $ccd$, respectively. According to the $SU(3)$ flavor symmetry, the DCBs with quark composition $ccs$, corresponding to $\Omega^{+}_{cc}$, should also exist, which is indeed newly observed by LHCb with the mass of $3725.9\pm 1.3$~MeV~\cite{lhcbconf}. 

The aforementioned exciting new measurements of $\Xi^{++}_{cc}$, $\Xi^{+}_{cc}$ and $\Omega^{+}_{cc}$, evidently complete the ground triplet states of the DCBs. Naturally the next step is to investigate the excited DCBs. We focus on the study of the possible excited DCBs resulting from the scattering of the ground DCBs and the light pseudoscalar mesons in this work. The combination of heavy quark symmetry~\cite{Manohar:2000dt,Savage:1990di} and chiral symmetry~\cite{Weinberg:1978kz,Gasser:1983yg,Gasser:1987rb} provides a reliable approach for this purpose.  

Within the framework of chiral perturbation theory ($\chi$PT), extensive theoretical studies have been conducted for the scattering processes between DCBs and the light pseudoscalar mesons $\pi, K, \eta$, that can be described as the pseudo-Nambu-Goldstone bosons (pNGBs) resulting from the spontaneous breaking of chiral symmetry. In Ref.~\cite{Guo:2017vcf}, the scattering of pNGBs with DCBs was investigated by taking the leading order (LO) $\xpt$ expressions in the unitarized approach, revealing possible signals of bound or resonant states in certain scattering channels. Using the chiral unitarization approach combined with heavy antiquark-diquark symmetry (HADS)~\cite{Savage:1990di}, Ref.~\cite{Yan:2018zdt} predicted several narrow negative-parity states with $J^P=1/2^-$. The $O(p^4)$ chiral effective Lagrangians describing the interactions between DCBs and pNGBs were constructed in Ref.~\cite{Qiu:2020omj}. Ref.~\cite{Liang:2023scp} further performed the one-loop calculation and predicted the $S$-wave and $P$-wave scattering lengths as well as the $S$-wave phase shifts for DCB–pNGB scattering, by estimating the values of the low-energy constants (LECs) using HADS. 

Although important progresses have been made in the study of DCBs within $\xpt$, there remains a lack of concrete constraints on the values of LECs relevant to the scatterings of DCBs and pNGBs. While the latter processes are still far beyond the reach of the current experiments, the lattice QCD can provide an alternative way to quantify these kinds of interactions. Based on the gauge ensembles generated by the CLQCD Collaboration~\cite{CLQCD:2023sdb}, the $S$-wave scattering lengths of the four elastic channels are calculated at unphysically large pion masses in Ref.~\cite{Yi:2025bnh}. One of the key novelties of this work is to exploit these valuable lattice data to constrain the values of the LECs appearing in the DCB-pNGB scattering processes within the framework of chiral unitarized approach by including the NLO amplitudes. Furthermore, we perform the chiral extrapolation to make predictions on the DCB-pNGB scattering at physical quark masses, including the scattering parameters and possible bound/virtual/resonance states.

In this work, we will concentrate on the $S$-wave scattering processes between the ground DCB states and the pNGBs within the framework of $SU(3)$ $\xpt$. We will construct the unitarized scattering amplitudes by taking the NLO $\xpt$ amplitudes with definite strangeness and isospin quantum numbers. By performing fits to the four elastic channels from lattice QCD simulations~\cite{CLQCD:2023sdb,Yi:2025bnh}, we determine the NLO LECs, which will be further used to predict various physical quantities appearing in the coupled channels.

The structure of this paper is organized as follows. In Sec.~\ref{sec.dcbamp}, we elaborate the theoretical framework for DCB–pNGB scattering based on $\xpt$, including the NLO chiral effective Lagrangian, the construction of scattering amplitudes with definite strangeness and isospin quantum numbers, the explicit expressions for $S$-wave partial wave projection, and the detailed procedure for implementing the unitarization. In Sec.~\ref{sec.rad}, we first perform fits to the lattice QCD simulation data based on the CLQCD ensembles to determine the NLO LECs. Meanwhile, based on the fitted parameters, we calculate the $S$-wave scattering lengths and phase shifts for various scattering channels, including both single and coupled channels. Finally, by analytically extrapolating the scattering amplitudes to the complex energy plane, we study the pole contents, including bound, virtual and resonance states, and analyze their coupling strengths to various scattering channels. Sec.~\ref{sec.sum} provides a summary and conclusions of our study.

\section{Chiral effective Lagrangian, partial-wave amplitude and its unitarization}\label{sec.dcbamp}

\subsection{Chiral effective Lagrangian up to NLO}

The physical states $\Xi^{++}_{cc}$,$\,\Xi^{+}_{cc}$ and $\Omega^{+}_{cc}$ constitute the DCB triplet
\begin{equation}
\begin{aligned}
\psi=
\begin{pmatrix}
\Xi^{++}_{cc}  \\
\Xi^{+}_{cc}  \\
\Omega^{+}_{cc} 
\end{pmatrix}\,.
\end{aligned} 
\end{equation}
The pseudoscalar mesons ($\pi$,\,$K$,\,$\eta$) are treated as the pNGBs resulting from spontaneous chiral symmetry breaking in QCD, which can be incorporated into the chiral Lagrangian via the exponential parametrization
\begin{equation}
\begin{aligned}
    U=e^{\frac{i\sqrt{2}\phi}{F}}\,,
\end{aligned} 
\end{equation}
where the $SU(3)$ matrix representation of the pNGB octet is given by
\begin{equation}
\begin{aligned}
\phi=
\begin{pmatrix}
\frac{1}{\sqrt{2}}\pi^0+\frac{1}{\sqrt{6}}\eta & \pi^+ & K^{+} \\
\pi^- & -\frac{1}{\sqrt{2}}\pi^0+\frac{1}{\sqrt{6}}\eta & K^0 \\
K^- & \bar{K}^0 & -\frac{2}{\sqrt{6}}\eta
\end{pmatrix}\,.
\end{aligned}
\end{equation}
Here, $F$ refers to the pion decay constant in the chiral limit.

The chiral effective Lagrangian describing the scattering processes between DCBs and pNGBs up to NLO, i.e., $O(p^2)$, is given by~\cite{Qiu:2020omj}
\begin{equation}
\begin{aligned}
\mathcal{L}_{\psi\phi}=\mathcal{L}^{(1)}_{\psi\phi}+\mathcal{L}^{(2)}_{\psi\phi}\,,
\end{aligned}
\end{equation}
with  
\begin{align}
    \mathcal{L}^{(1)}_{\psi\phi}=&\,\bar{\psi}(iD\!\!\!\!/-m)\psi+\frac{g}{2}\bar{\psi}u\!\!\!/\gamma_{5}\psi\,,\\
\mathcal{L}^{(2)}_{\psi\phi}=&\,b_{1}\bar{\psi}\langle\chi_
{+}\rangle\psi+b_{2}\bar{\psi}(\chi_{+}-\frac{1}{3}\langle \chi_{+}\rangle)\psi+b_{3}\bar{\psi} u^2\psi+b_{4}\bar{\psi}\langle u^2\rangle\psi+
\frac{b_5}{\overset{\circ}{M^2}}\bar{\psi}(\{u^{\mu},u^{\nu}\}D_{\mu\nu}+H.c.)\psi\nonumber\\
&+\frac{b_6}{\overset{\circ}{M^2}}\bar{\psi}(\langle u^{\mu}u^{\nu} \rangle D_{\mu\nu}+H.c.)\psi+ib_{7}\bar{\psi}[u^{\mu},u^{\nu}]\sigma_{\mu\nu}\psi\,,
\end{align}
where the superscripts (1, 2) denote the chiral orders and the basic chiral building items are defined as 
\begin{equation}
\begin{aligned}
    D_{\mu}=&\,\partial_{\mu}+\frac{1}{2}\left(u^{\dagger}\partial_{\mu}u+u\partial_{\mu}u^{\dagger}\right)\,, &u_{\mu}=&\,i\left(u^{\dagger}\partial_{\mu}u-u\partial_{\mu}u^{\dagger}\right)\,,&u=&e^{\frac{i\phi}{\sqrt{2}F}}\,,\\ 
     \chi_{\pm}=&\,u^{\dagger}\chi u^{\dagger}\pm u\chi^{\dagger}u\,,
    &D_{\mu\nu}=&\,\{ D_{\mu},D_{\nu} \}\,,&\sigma_{\mu\nu}=&\frac{i}{2}\left[\gamma_{\mu},\gamma_{\nu}\right]\,.
\end{aligned}
\end{equation}
Here, $\overset{\circ}{M}$ denotes the mass of the DCB in the chiral limit, which is estimated by taking the average of the masses of $\Xi_{cc}^{++},\Xi_{cc}^{+}$ and $\Omega_{cc}^{++}$. $g$ represents the LO axial-vector coupling constant, $b_{i=1,...,7}$ are the NLO LECs with units of GeV$^{-1}$ and $H.c.$ denotes Hermitian conjugation.

For the scattering process $\psi_{1}(p)\phi_{1}(q)\rightarrow\psi_{2}(p')\phi_{2}(q')$, the Lorentz-invariant amplitude can be decomposed as
\begin{align}\label{eq.ampfull}    
\mathcal{T}_{\psi_{1}\phi_{1}\rightarrow\psi_{2}(p)\phi_{2}(q)}=\bar{u}(p',\sigma')\bigg[A(s,t)+\frac{1}{2}(q\!\!\!/+q\!\!\!/')B(s,t)\bigg]u(p,\sigma)\,,
\end{align}
where $\psi_{1,2}$ represent the incoming and outgoing DCBs,  $\phi_{1,2}$ denote the incoming and outgoing pNGBs, and $\sigma^{(')}$ are the spins of the corresponding DCBs. 
$A(s,t)$ and $B(s,t)$ include the LO contact terms (usually named as Weinberg–Tomozawa (WT) term), the $s$-channel and $u$-channel contributions, as well as the NLO contact terms (ct). We explicitly write them as 
\begin{equation}
\begin{aligned}
    A(s,t)=A_{WT}(s,t)+A_{s}(s,t)+A_{u}(s,t)+A_{ct}(s,t)\,,\\
    B(s,t)=B_{WT}(s,t)+B_{s}(s\,t)+B_{u}(s,t)+B_{ct}(s,t)\,,
\end{aligned}
\end{equation}
where the first three terms for $A(s,t)$ and $B(s,t)$ are contributed by the LO Lagrangian and the last term is originated from the NLO one. The Mandelstam variables are defined in the usual way as $s=(p+q)^2,t=(p-p')^2$ and $u=(p-q')^2$, satisfying $s+t+u=m^2_{\psi_1}+m^2_{\phi_1}+m^2_{\psi_2}+m^2_{\phi_2}$. 

\subsection{Partial wave amplitude and its unitarization}

For the DCBs $(\Xi_{cc}^{++},\Xi_{cc}^{+},\Omega_{cc}^{+})$ and pNGBs $(\pi,K,\eta)$ scattering processes, it is customary to classify different reactions according to the quantum numbers of $(S\,,I)$, with strangeness $S$ and isospin $I$. For the DCBs and pNGBs in the isospin basis $|I,I_{3}\rangle$, we adopt the following conventions 
\begin{equation}
\begin{aligned}
|\Omega_{cc}^{+}\rangle&=|0,0\rangle\,,&|\bar{K^0}\rangle&=-|\frac{1}{2},\frac{1}{2}\rangle,&|K^{-}\rangle&=|\frac{1}{2},-\frac{1}{2}\rangle\,,
\end{aligned}
\end{equation}
for particles with $S=-1$, 
\begin{equation}
\begin{aligned}
    |\Xi_{cc}^{++}\rangle&=|\frac{1}{2},\frac{1}{2}\rangle\,,&|\Xi_{cc}^{+}\rangle&=|\frac{1}{2},-\frac{1}{2}\rangle\,,&|\eta\rangle&=|0,0\rangle\,,\\
    |\pi^{+}\rangle&=-|1,1\rangle\,,&|\pi^{0}\rangle&=|1,0\rangle\,,&|\pi^{-}\rangle&=|1,-1\rangle\,,
\end{aligned}
\end{equation}
for particles with $S=0$, and
\begin{equation}
\begin{aligned}
    |K^{+}\rangle&=|\frac{1}{2},\frac{1}{2}\rangle\,,&|K^0\rangle&=|\frac{1}{2},-\frac{1}{2}\rangle\,,
\end{aligned}
\end{equation}
for those with $S=1$. 
The processes $\psi_{1}\phi_{1}\rightarrow\psi_{2}\phi_{2}$ include four elastic scattering cases and three cases with coupled channels. The relations between the physical scattering amplitudes and those with definite quantum numbers $(S,I)$ are given as follows.
For the elastic channel with $(S,I)=(-2,\frac{1}{2})$, it is given by 
\begin{equation}
\begin{aligned}
\mathcal{T}^{(-2\,,\frac{1}{2})}_{\Omega_{cc}\bar{K}\rightarrow\Omega_{cc}\bar{K}}(s,t,u)&=\mathcal{T}_{\Omega^{+}_{cc}K^{-}\rightarrow\Omega^{+}_{cc}K^{-}}(s,t,u)\,. 
\end{aligned}
\end{equation}
For the elastic channel with $(S,I)=(1,1)$, it reads 
\begin{equation}
\begin{aligned}
\mathcal{T}^{(1,1)}_{\Xi_{cc}K\rightarrow\Xi_{cc}K}(s,t,u)&=\mathcal{T}_{\Xi^{++}_{cc}K^{+}\rightarrow\Xi^{++}_{cc}K^{+}}(s,t,u)\,. 
\end{aligned}
\end{equation}
For the elastic channel with $(S,I)=(1,0)$, it takes the form  
\begin{equation}
\begin{aligned}
\mathcal{T}^{(1,0)}_{\Xi_{cc}K\rightarrow\Xi_{cc}K}(s,t,u)&=2\mathcal{T}_{\Xi^{++}_{cc}K^{0}\rightarrow\Xi^{++}_{cc}K^{0}}(s,t,u)-\mathcal{T}_{\Xi^{++}_{cc}K^{+}\rightarrow\Xi^{++}_{cc}K^{+}}(s,t,u)\,. 
\end{aligned}
\end{equation}
For the elastic channel with $(S,I)=(0,\frac{3}{2})$, it is 
\begin{equation}
\begin{aligned}
\mathcal{T}^{(0,\frac{3}{2})}_{\Xi_{cc}\pi\rightarrow\Xi_{cc}\pi}(s,t,u)&=\mathcal{T}_{\Xi^{++}_{cc}\pi^{+}\rightarrow\Xi^{++}_{cc}\pi^{+}}(s,t,u)\,. 
\end{aligned}
\end{equation}
The processes with $(S,I)=(-1,1)$ include two channels and their amplitudes read 
\begin{align}
\mathcal{T}^{(-1,1)}_{\Omega_{cc}\pi\rightarrow\Omega_{cc}\pi}(s,t,u)&=\mathcal{T}_{\Omega^{+}_{cc}\pi^{0}\rightarrow\Xi^{+}_{cc}\pi^{0}}(s,t,u)\,,\\
\mathcal{T}^{(-1,1)}_{\Xi_{cc}\bar{K}\rightarrow\Xi_{cc}\bar{K}}(s,t,u)&=\mathcal{T}_{\Xi^{++}_{cc}K^{0}\rightarrow\Xi^{++}_{cc}K^{0}}(u,s,t)\,, \\
\mathcal{T}^{(-1,1)}_{\Omega_{cc}\pi\rightarrow\Xi_{cc}\bar{K}}(s,t,u)&=\sqrt{2}\mathcal{T}_{\Xi^{++}_{cc}K^{-}\rightarrow\Omega^{+}_{cc}\pi^{0}}(s,t,u)\,.
\end{align}
The scattering processes with $(S,I)=(-1,0)$ also include two channels and the amplitudes are  
\begin{align}
\mathcal{T}^{(-1,0)}_{\Xi_{cc}\bar{K}\rightarrow\Xi_{cc}\bar{K}}(s,t,u)&=2\mathcal{T}_{\Xi^{++}_{cc}K^{+}\rightarrow\Xi^{++}_{cc}K^{+}}(u,s,t)-\mathcal{T}_{\Xi^{++}_{cc}K^{0}\rightarrow\Xi^{++}_{cc}K^{0}}(u,s,t)\,,  \\
\mathcal{T}^{(-1,0)}_{\Omega_{cc}\eta\rightarrow\Omega_{cc}\eta}(s,t,u)&=\mathcal{T}_{\Omega^{+}_{cc}\eta\rightarrow\Omega^{+}_{cc}\eta}(s,t,u)\,,\\
\mathcal{T}^{(-1,0)}_{\Xi_{cc}\bar{K}\rightarrow\Omega_{cc}\eta}(s,t,u)&=\sqrt{2}\mathcal{T}_{\Xi^{+}_{cc}\bar{K}^{0}\rightarrow\Omega^{+}_{cc}\eta}(s,t,u)\,.
\end{align}
The processes with $(S,I)=(0,\frac{1}{2})$ have three coupled channels and they take the forms 
\begin{align}
\mathcal{T}^{(0,\frac{1}{2})}_{\Xi_{cc}\pi\rightarrow\Xi_{cc}\pi}(s,t,u)&=\frac{3}{2}\mathcal{T}_{\Xi^{++}_{cc}\pi^{+}\rightarrow\Xi^{++}_{cc}\pi^{+}}(u,s,t)-\frac{1}{2}\mathcal{T}_{\Xi^{++}_{cc}\pi^{+}\rightarrow\Xi^{++}_{cc}\pi^{+}}(s,t,u)\,, \\
\mathcal{T}^{(0,\frac{1}{2})}_{\Xi_{cc}\eta\rightarrow\Xi_{cc}\eta}(s,t,u)&=\mathcal{T}_{\Xi^{++}_{cc}\eta\rightarrow\Xi^{++}_{cc}\eta}(s,t,u)\,, \\
\mathcal{T}^{(0,\frac{1}{2})}_{\Omega_{cc}K\rightarrow\Omega_{cc}K}(s,t,u)&=\mathcal{T}_{\Omega^{+}_{cc}K^{-}\rightarrow\Omega^{+}_{cc}K^{-}}(u,s,t)\,, \\
\mathcal{T}^{(0,\frac{1}{2})}_{\Xi_{cc}\pi\rightarrow\Xi_{cc}\eta}(s,t,u)&=\sqrt{3}\mathcal{T}_{\Xi^{++}_{cc}\pi^{0}\rightarrow\Xi^{++}_{cc}\eta}(s,t,u)\,, \\
\mathcal{T}^{(0,\frac{1}{2})}_{\Xi_{cc}\pi\rightarrow\Omega_{cc}K}(s,t,u)&=\sqrt{3}\mathcal{T}_{\Xi^{++}_{cc}K^{-}\rightarrow\Omega^{+}_{cc}\pi^{0}}(u,s,t)\,, \\
\mathcal{T}^{(0,\frac{1}{2})}_{\Xi_{cc}\eta\rightarrow\Omega_{cc}K}(s,t,u)&=\mathcal{T}_{\Xi^{+}_{cc}\bar{K}^{0}\rightarrow\Omega^{+}_{cc}\eta}(u,s,t)\,.
\end{align}

Next we perform the partial-wave projection of the full amplitudes in Eq.~\eqref{eq.ampfull}. 
We focus on the $S$-wave case, i.e., by taking $l=0$, and the corresponding partial-wave projection formula reads~\cite{Guo:2017vcf}
\begin{equation}\label{eq.pwt}
\begin{aligned}
    \mathcal{T}^{(S,I)}_{l=0}(s)=\sum_{\sigma,\sigma'}\frac{1}{8\pi}\int d\Omega \,\mathcal{T}(s,\Omega;\sigma,\sigma')\,,
\end{aligned}
\end{equation}
where $\Omega$ represents the solid angle of the three-momentum of the final state in the center of mass (CM) frame. The explicit $S$-wave expressions for the DCB-pNGB scattering amplitudes with ${(S,I)}$ are found to be 
\begin{equation}\label{eq.yy1}
\begin{aligned}
    \mathcal{T}^{(S,I)}_{l=0}(s)=&\frac{1}{\sqrt{E_{\psi_{1}}+m_{\psi_{1}}}\sqrt{E_{\psi_{2}}+m_{\psi_{2}}}}\Bigg\{A_{s,0}^{(S,I)}(s)+\frac{2\sqrt{s}-m_{\psi_{1}}-m_{\psi_{2}}}{2}\\
   & \times\left[B_{WT,0}^{(S,I)}(s)+B_{s,0}^{(S,I)}(s)\right] +\Bigg[\Bigg(1-\frac{|\vec{p}_{\psi_{1}}||\vec{p}_{\psi_{2}}|}{(E_{\psi_{1}}+m_{\psi_{1}})(E_{\psi_{2}}+m_{\psi_{2}})}\Bigg)\\
   &\left.\left.\times(A_{u,1}^{(S,I)}(s)+A_{ct,1}^{(S,I)}(s)\right)\Bigg]+\frac{1}{2}\Bigg[(2\sqrt{s}-m_{\psi_{1}}-m_{\psi_{2}})\left(B_{u,0}^{(S,I)}(s)\right.\right.\\
   &\left.+B_{ct,0}^{(S,I)}(s)\right)+\frac{(2\sqrt{s}-m_{\psi_{1}}-m_{\psi_{2}})|\vec{p}_{\psi_{1}|}|\vec{p}_{\psi_{2}}|}{(E_{\psi_{1}}+m_{\psi_{1}})(E_{\psi_{2}}+m_{\psi_{2}})}B^{(S,I)}_{ct,1}(s)\Bigg]
    \Bigg\}\,, 
\end{aligned}
\end{equation}
where $E_{\psi_{1}}$ and $E_{\psi_{2}}$ are the CM energies of the incoming and outgoing DCBs, respectively, and $\vec{p}_{\psi_{1}}$ and $\vec{p}_{\psi_{2}}$ are their CM three-momenta, which are given by
\begin{align}
    E_{\psi_{1,2}}&=\frac{s+m_{\psi_{1,2}}-m_{\phi_{1,2}}}{2\sqrt{s}}\,,\\
    p_{\psi_{1,2}}&=|\vec{p}_{\psi_{1,2}}|=\frac{\sqrt{\left[s-(m_{\psi_{1,2}}+m_{\phi_{1,2}})^2\right]\left[s-(m_{\psi_{1,2}}-m_{\phi_{1,2}})^2\right]}}{2\sqrt{s}}\,.
\end{align}
Both the LO and NLO contributions are included in $A^{(S,I)}_{j,0(1)}(s)$ and $B^{(S,I)}_{j,0(1)}(s)$ of Eq.\eqref{eq.yy1}. The $WT$ term and the $s$- and $u$-exchanges correspond to the LO parts. The explicit expressions for the LO parts of $A^{(S,I)}_{j,0(1)}(s)$ and $B^{(S,I)}_{j,0(1)}(s)$ are given by 
\begin{align}
    A^{(S,I)}_{s,0}(s)=&\frac{C_{s}g^2}{F_{\phi_1}F_{\phi_2}}\frac{m_{ex}(m_{\psi_1}^2+m_{\psi_2}^2-2s)+(m_{\psi_1}+m_{\psi_2})(m_{\psi_1}+m_{\psi_2}-2s)}{m_{ex}^2-s}\,,\\    
    A^{(S,I)}_{u,0}(s)=&\frac{C_{u}g^2}{F_{\phi_1}F_{\phi_2}}\bigg\{ 2m_{ex}[2E_ {\psi_1}E_{\psi_2}+m_{\phi_1}^2+m_{\phi_2}^2-2(m_{\psi_{1}}^2+m_{\psi_{2}}^2+s)+\Sigma]\nonumber\\
    &+\frac{(m_{\psi_1}+m_{\psi_2})(m_{\psi_1}m_{\psi_{2}}-\Sigma)}{2|\vec{p}_{\psi_1}||\vec{p}_{\psi_2}|}\log\frac{F_{+}(s)}{F_{-}(s)}\bigg\}\,,\\
    A^{(S,I)}_{u,1}(s)=&\frac{C_{u}g^2}{3F_{\phi_1}F_{\phi_2}}\bigg [-8m_{ex}|\vec{p}_{\psi_1}||\vec{p_{\psi_2}}|+\frac{3(m_{\psi_1}+m_{\psi_2})(m_{\psi_1}m_{\psi_2}-\Sigma)}{|\vec{p}_{\psi_1}||\vec{p}_{\psi_2}|}+\frac{(m_{\psi_1}+m_{\psi_2})}{4|\vec{p}_{\psi_1}|^2|\vec{p}_{\psi_2}|^2}\nonumber\\
    &\times(m_{\psi_1}m_{\psi_2}-\Sigma)(m_{ex}^2+m_{\psi_1}^2+m_{\psi_2}^2+s-\Sigma-E_{\psi_1}E_{\psi_2})\log\frac{F_{-}(s)}{F_{+}(s)}\bigg]\,,\\
    B^{(S,I)}_{WT,0}(s)=&\frac{C_{WT}}{F_{\phi_1}F_{\phi_2}}\,,\\
    B^{(S,I)}_{s,0}(s)=&\frac{2C_{s}g^2}{F_{\phi_1}F_{\phi_2}}\frac{m_{\psi_{1}}m_{\psi_{2}}+m_{ex}(m_{\psi_1}+m_{\psi_2})+s}{m_{ex}^2-s}\,,\\
    B^{(S,I)}_{u,0}(s)=&\frac{C_{u}g^2}{F_{\phi_1}F_{\phi_2}}\left\{4+\frac{1}{|\vec{p}_{\psi_1}||\vec{p}_{\psi_2}|}(m_{ex}+m_{\psi_{1}})(m_{ex}+m_{\psi_{2}})\log\frac{F_{-}(s)}{F_{+}(s)}\right\}\,,\\
    B^{(S,I)}_{u,1}(s)=&-\frac{C_{u}g^2}{F_{\phi_1}F_{\phi_2}}\bigg\{\frac{(m_{ex}+m_{\psi_1})(m_{ex}+m_{\psi_2})}{2|\vec{p}_{\psi_1}|^2|\vec{p}_{\psi_{2}}|^2}\bigg[4|\vec{p}_{\psi_1}||\vec{p}_{\psi_2}|+(m_{ex}^2+m_{\psi_1}^2\nonumber\\&+m_{\psi_2}^2-E_{\psi_1}E_{\psi_2}
    +s-\Sigma)\log\frac{F_{+}(s)}{F_{-}(s)}\bigg]\bigg\}\,,
\end{align}
where $\Sigma=m_{\psi_{1}}^2+m_{\phi_{1}}^2+m_{\psi_{2}}^2+m_{\phi_{2}}^2$, $m_{ex}$ denotes the physical mass of the exchanged DCB and the functions $F_{\pm}(s)$ are defined as
\begin{equation}
\begin{aligned}
    F_{\pm}(s)&=m_{ex}^2+m_{\psi_1}^2+m_{\psi_2}^2-E_{\psi_{1}}E_{\psi_{2}}+s-\Sigma\pm2|\vec{p}_{\psi_1}||\vec{p}_{\psi_2}|\,.
\end{aligned}
\end{equation}
The coefficients $C_{WT}$, $C_{s}$ and $C_{u}$, corresponding to the Weinberg–Tomozawa term, the $s$-channel, and the $u$-channel at LO, respectively, are collected in Table~\ref{tab.cicoef} for each channel. It is noted that we distinguish the pNGB decay constants $F_{\phi}$ for each channel, namely $F_\pi$, $F_K$ and $F_\eta$ will be used for the channels involving $\pi, K$ and $\eta$ states, respectively.   

The NLO contributions only include the contact terms and their explicit expressions read 
\begin{align}
    A^{(S,I)}_{ct,0}(s)=&\frac{C_1}{F_{\phi_1}F_{\phi_2}}+\frac{C_2}{F_{\phi_1}F_{\phi_2}}(E_{\psi_1}E_{\psi_2}+m_{\phi_1}^2+m_{\phi_2}^2-m_{\psi_1}^2-m_{\psi_2}^2)+\frac{C_3}{\overset{\circ}{M} F_{\phi_1}F_{\phi_2}}[m_{\psi_1}^4+m_{\psi_2}^4\nonumber\\    &+m_{\phi_2}^2(m_{\psi_1}^2+m_{\psi_2}^2)+m_{\phi_1}^2(2m_{\phi_2}^2+m_{\psi_1}^2+m_{\psi_2}^2)+2E_{\psi_1}E_{\psi_2}s-2m_{\psi_1}^2s\nonumber\\
    &-2m_{\psi_2}^2s-2s^2+(m_{\psi_1}^2+m_{\psi_2}^2+2s-2E_{\psi_1}E_{\psi_2})\Sigma-\Sigma^2]+\frac{C_4}{F_{\phi_1}F_{\phi_2}}(E_{\psi_1}E_{\psi_2}\nonumber\\
    &-m_{\psi_1}^2-m_{\psi_2}^2-2s-2\Sigma)\,,\\
    A^{(S.I)}_{ct,1}(s)=&\frac{2C_3|\vec{p}_{\psi_1}||\vec{p}_{\psi_2}|}{3\overset{\circ}{M}F_{\phi_1}F_{\phi_2}}(\Sigma-2s)-\frac{2(C_2+C_4)|\vec{p}_{\psi_1}||\vec{p}_{\psi_2}|}{3F_{\phi_1}F_{\phi_2}}\,,\\
    B^{(S,I)}_{ct,0}(s)=&\frac{2C_4}{F_{\phi_1}F_{\phi_2}}(m_{\psi_1}+m_{\psi_2})\,,
\end{align}
where the quantities of $C_1$, $C_2$, $C_3$, $C_4$ for different channels are collected in Table~\ref{tab.cicoef}. The expressions of the LO and NLO coefficients have been provided in Refs.~\cite{Guo:2017vcf,Liang:2023scp} and they are explicitly given here for the sake of completeness.

\begin{table} [ht] 
\scriptsize
\renewcommand{\arraystretch}{2.0}
\begin{tabular}{c c|p{0.4cm} p{1.1cm} p{1.1cm} p{2.7cm} p{1.55cm} p{1.55cm} c} 
\hline
 \((S,I)\) & Processes & \centering \(C_{WT}\) &\centering \(C_{s}\)& \centering \(C_{u}\)& \centering  \(C_{1}\) & \centering \(C_{2}\) & \centering \(C_{3}\)& 
\(\centering C_{4}\) \\
\hline 
\centering \((-2,\frac{1}{2})\) & \(\Omega_{cc}\bar{K}\rightarrow\Omega_{cc}\bar{K}\) & \(\centering -1\)& \centering 0 & \centering \(\frac{1}{4}[\Xi_{cc}]\)& \centering \(-\frac{4}{3}(6b_1+b_2)m^2_{K}\) & \centering \(2(b_3+2b_4)\) & \centering \(4(b_5+b_6)\)& \(-4b_7\) \\
\hline 
\centering \((1,1)\) & \(\Xi_{cc}K\rightarrow\Xi_{cc}K\) & \centering \(-1\) & \centering 0 & \centering \(\frac{1}{4}[\Omega_{cc}]\) & \centering \(-\frac{4}{3}(6b_1+b_2)m^2_{K}\) & \centering  \(2(b_3+2b_4)\) & \centering  \(4(b_5+b_6)\)&\(-4b_7\) \\
\hline  
\((1,0)\) & \(\Xi_{cc}K\rightarrow\Xi_{cc}K\) & \centering  \(1\) & \centering 0 &\centering  \(-\frac{1}{4}[\Omega_{cc}]\) &\centering \(-\frac{4}{3}(6b_1-5b_2)m^2_{K}\) &\centering \(-2(b_3-2b_4)\)& \centering \(-4(b_5-b_6)\)&\(4b_7\) \\
\hline
\((0,\frac{3}{2})\) & \(\Xi_{cc}\pi\rightarrow\Xi_{cc}\pi\) &\centering\(-1\)& \centering  0 & \centering \(\frac{1}{4}[\Xi_{cc}]\)&\centering \(-\frac{4}{3}(6b_1+b_2)m^2_{\pi}\) & \centering \(2(b_3+2b_4)\) & \centering \(4(b_5+b_6)\)&\(-4b_7\) \\
\hline 
\((-1,0)\) & \(\Xi_{cc}\bar{K}\rightarrow\Xi_{cc}\bar{K}\) &\centering \(2\)& \centering\(1[\Omega_{cc}]\)&\centering 0& \centering \(-\frac{8}{3}(3b_1+2b_2)m^2_{K}\) &\centering \(4(b_3+b_4)\) &\centering \(4(2b_5+b_6)\)&\(8b_7\) \\
  & \(\Omega_{cc}\eta\rightarrow\Omega_{cc}\eta\) &\centering 0&\centering \(\frac{1}{3}[\Omega_{cc}]\)&\centering \(\frac{1}{6}[\Omega_{cc}]\)&\centering  \(-\frac{32}{9}(3b_1+2b_2)m^2_{K}+\frac{8}{9}(3b_1+5b_2)m^2_{\pi}\) &\centering  \(\frac{4}{3}(2b_3+3b_4)\) & \centering \(\frac{4}{3}(4b_5+3b_6)\)& 0 \\ 
  & \(\Xi_{cc}\bar{K}\rightarrow\Omega_{cc}\eta\) &\centering\(-\sqrt{3}\)&\centering \(-\frac{1}{\sqrt{3}}[\Omega_{cc}]\)& \centering\(-\frac{1}{4\sqrt{3}}[\Xi_{cc}]\)&\centering\(\frac{2\sqrt{3}}{3}b_2(5m^2_{K}-3m^2_{\pi})\) &\centering \(-\frac{2}{\sqrt{3}}b_3\) & \centering\(-\frac{4}{\sqrt{3}}b_5\) & \(-4\sqrt{3}b_7\) \\
\hline 
\((-1,1)\) & \(\Omega_{cc}\pi\rightarrow\Omega_{cc}\pi\) &\centering 0&\centering 0&\centering 0 &\centering  \(-\frac{8}{3}(3b_1-b_2)m^2_{\pi}\) & \centering \(4b_4\) &\centering  \(4b_6\) & 0 \\
  & \(\Xi_{cc}\bar{K}\rightarrow\Xi_{cc}\bar{K}\) &\centering 0&\centering 0&\centering 0& \centering \(-\frac{8}{3}(3b_1-b_2)m^2_{K}\) & \centering \(4b_4\) & \centering  \(4b_6\) & 0\\
  & \(\Omega_{cc}\pi\rightarrow\Xi_{cc}\bar{K}\) &\centering \(-1\)&\centering 0&\centering \(\frac{1}{4}[\Xi_{cc}]\)&\centering  \(-2b_2(m^2_{K}+m^2_{\pi})\) & \centering \(2b_3\) &\centering  \(4b_5\) & \(-4b_7\)  \\
\hline 
\((0,\frac{1}{2})\) & \(\Xi_{cc}\pi\rightarrow\Xi_{cc}\pi\) & \centering 2&\centering \(\frac{3}{4}[\Xi_{cc}]\)&\centering  \(-\frac{1}{8}[\Xi_{cc}]\) &\centering \(-\frac{4}{3}(6b_1+b_2)m^2_{\pi}\) &\centering  \(2(b_3+2b_4)\) & \centering \(4(b_5+b_6)\) & \(8b_7\)\\
  & \(\Xi_{cc}\eta\rightarrow\Xi_{cc}\eta\) &\centering 0&\centering \(\frac{1}{12}[\Xi_{cc}]\)&\centering \(\frac{1}{24}[\Xi_{cc}]\)& \centering \(-\frac{32}{9}(3b_1-b_2)m^2_{K}+\frac{8}{9}(3b_1-5b_2)m^2_{\pi}\) &\centering \(\frac{2}{3}(b_3+6b_4)\) &\centering  \(\frac{4}{3}(b_5+3b_6)\) & 0\\
  & \(\Omega_{cc}K\rightarrow\Omega_{cc}K\) &\centering \(1\)&\centering \(\frac{1}{2}[\Xi_{cc}]\)&\centering 0&\centering  \(-\frac{4}{3}(6b_1+b_2)m^2_{K}\) & \centering\(2(b_3+2b_4)\) & \centering \(4(b_5+b_6)\) & \(4b_7\) \\
  & \(\Xi_{cc}\pi\rightarrow\Xi_{cc}\eta\) &\centering 0&\centering \(\frac{1}{4}[\Xi_{cc}]\)&\centering \(\frac{1}{8}[\Xi_{cc}]\)&\centering  \(-4b_2m^2_{\pi}\) &\centering  \(2b_3\) &\centering \(4b_5\) & 0\\
  & \(\Xi_{cc}\pi\rightarrow\Omega_{cc}K\) & \centering \(\frac{\sqrt{6}}{2}\)&\centering \(\frac{\sqrt{6}}{4}[\Xi_{cc}]\)& \centering 0&\centering \(-\sqrt{6}b_2(m^2_{K}+m^2_{\pi})\) &\centering  \(\sqrt{6}b_3\) &\centering  \(2\sqrt{6}b_5\) & \(2\sqrt{6}b_7\)\\
  & \(\Xi_{cc}\eta\rightarrow\Omega_{cc}K\) &\centering \(\frac{\sqrt{6}}{2}\)&\centering \(\frac{\sqrt{6}}{12}[\Xi_{cc}]\)&\centering \(-\frac{\sqrt{6}}{12}[\Omega_{cc}]\)&\centering  \(\frac{\sqrt{6}}{3}b_2(5m^2_{K}-3m^2_{\pi})\) &\centering  \(-\frac{\sqrt{6}}{3}b_3\) &\centering  \(-\frac{2\sqrt{6}}{3}b_5\) & \(2\sqrt{6}b_7\) \\
\hline  
\end{tabular}
\centering
\caption{Coefficients for the LO and NLO amplitudes. The intermediate DCBs exchanged in the $s$ and $u$ channels are indicated inside brackets in the columns of $C_{s}$ and $C_u$. }\label{tab.cicoef}  
\end{table}  

It is possible that bound or resonant states could arise from the scattering processes between DCBs and pNGBs. However, the perturbative $\xpt$ partial-wave amplitudes in Eq.~\eqref{eq.pwt} are insufficient to study these interesting bound or resonant states.  A reliable way to investigate the latter states is to perform the unitarization by taking the perturbative $\xpt$ amplitudes as inputs. 
Unitarity imposes a fundamental constraint in this procedure. In this work, we adopt the unitarized chiral perturbation theory (UChPT) framework from Ref.~\cite{Oller:1998zr} to construct scattering amplitudes as
\begin{equation}\label{eq.unit}
\begin{aligned}
      T^{\mathrm{UChPT}}(s)=\frac{\mathcal{T}^{(S,I)}_{l=0}(s)}{1-\mathcal{T}^{(S,I)}_{l=0}(s)\cdot G(s)}\,,
\end{aligned}
\end{equation}
where $\mathcal{T}_{l=0}^{(S,I)}$ denotes the perturbative $S$-wave amplitudes up to NLO and $G(s)$ is the standard two-point one-loop function, which incorporates the contribution from the right-hand cut (RHC) associated with intermediate DCB and pNGB states. In the following the subscript of $l=0$ in the scattering amplitudes will be omitted for simplicity. 
The explicit integral representation of $G(s)$ is
\begin{equation}
\begin{aligned}
    G(s)=-i\int\frac{\mathrm{d}^4q}{(2\pi)^4}\frac{1}{(q^2-m_{1}^2+i\epsilon)[(P-q)^2-m^2_{2}+i\epsilon]}\,,
\end{aligned}
\end{equation}
where $m_1$ and $m_2$ stand for the masses of the two intermediate states and $s=P^2$. This loop function contains an ultraviolet divergence, which we regularize by adopting a sharp momentum cutoff scheme. After integrating over the energy component $q_0$, it simplifies to
\begin{equation}
\begin{aligned}
    G^{\Lambda}(s)=-\int^{|\vec{q}|<\Lambda}\frac{\mathrm{d}^3\vec{q}}{(2\pi)^3}\frac{\sqrt{|\vec{q}|^2+m_1^2}+\sqrt{|\vec{q}|^2+m_2^2}}{2\sqrt{(|\vec{q}|^2+m_1^2)(|\vec{q}|^2+m_2^2)}\big[s-(\sqrt{|\vec{q}|^2+m_1^2}+\sqrt{|\vec{q}|^2+m_2^2})^2+i\epsilon\big]}\,,
\end{aligned}
\end{equation}
where $\Lambda$ denotes the three-momentum cutoff and will be fitted in later discussion. The imaginary part of the $G(s)$ is 
\begin{equation}
 {\rm Im} G(s) = \rho(s)\,,
\end{equation}
with 
\begin{equation}
\begin{aligned}
    \rho(s)=\frac{p(s)}{8\pi\sqrt{s}}=\frac{\sqrt{[s-(m_{\psi_{1}}-m_{\phi_{1}})^2][s-(m_{\psi_1}+m_{\phi_{1}})^2]}}{16\pi s}\,,
\end{aligned}
\end{equation}
and $p(s)$ the CM momentum of the scattering particles. Since the perturbative amplitude $\mathcal{T}(s)$ up to NLO are real above the two-body threshold, it is easy to verify that the unitarized amplitude in Eq.~\eqref{eq.unit} satisfy the unitarity relation ${\rm Im}{T_{\rm UChPT}}^{-1}=\rho(s)$. For the coupled-channel scattering, the unitarized amplitude in Eq.~\eqref{eq.unit} should be extended to the matrix spanned in the reaction channels. 

Physical observables of the scattering process, such as the phase shift and the inelasticity coefficient, can be obtained from the $S$ matrix. The relation between the $S$ matrix and the unitarized partial-wave amplitude $T^{\text{UChPT}}$ in Eq.\eqref{eq.unit} is given by
\begin{equation}
\begin{aligned}
     S(s)=1+2i\sqrt{\rho(s)}\cdot T^{\text{UChPT}}(s) \cdot \sqrt{\rho(s)}\,,
\end{aligned}
\end{equation}
which satisfies the unitarity relation $S^{\dagger}S=1$. 
For the coupled-channel case, the diagonal phase shifts $\delta_{kk}$ and inelasticity parameters $\epsilon_{kk}$, as well as the off‑diagonal $(k\neq j)$ phase shifts $\delta_{kj}$ and inelasticity parameters $\epsilon_{kj}$, can be extracted from the $S$-matrix elements via 
\begin{equation}
\begin{aligned}
    S_{kk}=\epsilon_{kk}e^{2i^{\delta_{kk}}}\,,&\quad \quad S_{kj}=i\epsilon_{kj}e^{i\delta_{kj}}\,.
\end{aligned}
\end{equation}
The inelasticity parameters are obtained from the moduli of the corresponding $S$-matrix elements, $\epsilon_{kk(kl)}=|S_{kk(kj)}|$, and satisfy the relationship $0\leq\epsilon_{kk(kj)}\leq1$.

\section{Numerical results and phenomenological discussions}\label{sec.rad}

\subsection{Fits to the lattice data}

The lattice QCD simulations on the four single-channel $S$-wave scattering processes, including $\Omega_{cc}\bar{K}\,(-2,\frac{1}{2})$, $\Xi_{cc}K\,(1,0)$, $\Xi_{cc}K\,(1,1)$ and $\Xi_{cc}\pi\,(0,\frac{3}{2})$, where the numbers inside the brackets correspond to the quantum numbers of $(S,I)$, are performed in Refs.~\cite{Yi:2025bnh,Liang:2025kkn}. These data consist of four sets of 2+1 flavor ensembles with different volumes, corresponding to two different pion masses at approximately $210~\text{MeV}$ and $300~\text{MeV}$. 
The unphysical masses of the pion, kaon and DCBs resulting from those lattice ensembles~\cite{Yi:2025bnh,Liang:2025kkn} that will be used in our fits to the lattice data are summarized in Table~\ref{tab.latensem}. For the decay constants of pion and kaon at unphysical meson masses, we use the chiral extrapolation formulas provided in Ref.~\cite{Gao:2022xqz} that are obtained by fitting large amount of relevant lattice data based on the NLO $U(3)$ $\xpt$. The explicit values of $F_\pi$ and $F_K$ used in later fits for each ensemble are indicated in the last two columns of  Table~\ref{tab.latensem}.

\begin{table} [htbp] 
\centering
\renewcommand{\arraystretch}{1.5}
\begin{tabular}{c c c c c c c}  
\hline 
Ensembles & \(m_{\pi}\)(MeV) & \(m_{K}\)(MeV) & \(m_{\Xi_{cc}}\)(MeV) & \(m_{\Omega_{cc}}\)(MeV) & \(F_{\pi}\)(MeV) & \(F_{K}\)(MeV)\\
\hline 
F32P30 & 303.96 & 523.00 & 3633.02 & 3713.93 & 97.86 & 111.86 \\
F48P30 & 304.87 & 524.07 & 3636.90 & 3717.90 & 97.90 & 111.88 \\
F32P21 & 208.50 & 491.73 & 3605.50 & 3693.80 & 94.25 & 110.00 \\
F48P21 & 207.74 & 491.09 & 3608.00 & 3699.50 & 94.23 & 109.99  \\
\hline 
\end{tabular}
\caption{ Different masses from various ensembles from lattice QCD simulations are taken from Refs.~\cite{Yi:2025bnh,Liang:2025kkn}. The pion-mass depependences of $F_{\pi}$ and $F_{K}$ are calculated from Ref.~\cite{Gao:2022xqz}.}\label{tab.latensem}  
\end{table}

For the physical masses of the DCBs, $\Xi_{cc}^{+}$ and $\Xi_{cc}^{++}$ form an isospin doublet and are treated as degenerate in the isospin‑symmetric limit. Their common mass is taken as $m_{\Xi_{cc}}=3621.6$~MeV, according to the Particle Data Group (PDG)~\cite{ParticleDataGroup:2024cfk}. For $\Omega_{cc}^{+}$, we use the recent LHCb measurement~\cite{lhcbconf} with $m_{\Omega_{cc}}=3725.9$~MeV. The physical masses of the pNGBs are taken from the PDG: $m_{\pi}=139.57$~MeV, $m_{K}=493.68$~MeV and $m_{\eta}=547.86$~MeV. The physical decay constants of the $\pi$ and $K$ are set to their PDG average values with $F_{\pi}=92.2$~MeV and $F_{K}=112.0$~MeV. 
For the $\eta$ meson, the physical $\eta$ and $\eta'$ arise from a mixture of the singlet $\eta_{0}$ and the octet $\eta_{8}$. In $SU(3)$ chiral theory, the singlet $\eta_{0}$ is not included and the physical $\eta$ is given by the octet state $\eta_{8}$. Therefore, we treat the octet $\eta_{8}$ as an approximation of the physical $\eta$ particle. Accordingly, we use the $\eta$ decay constant obtained from $SU(3)$ $\xpt$, which leads to $F_{\eta} \simeq 117\,\text{MeV}$, according to Ref.~\cite{GomezNicola:2001as}. 

\begin{table} [htbp] 
\centering
\renewcommand{\arraystretch}{1.4}
\small
\begin{tabular}{c c c c c c c} 
\hline 
\((S,I)\) & Processes & \(m_{\pi}\)(MeV) & \(a_{0}\)(UChPT) & \(r_0\)(UChPT)&\(a_{0}\)(Ref.\cite{Yi:2025bnh}) & \(r_{0}\)(Ref.\cite{Yi:2025bnh})  \\
\hline
\((-2,\frac{1}{2})\) & \(\Omega_{cc}\bar{K}\rightarrow\Omega_{cc}\bar{K}\) & 210 & \(-0.168^{+0.014}_{-0.014}\) &\(0.05^{+0.08}_{-0.08}\) & \(-0.136\pm0.012\) &\(-0.67\pm0.25\)\\
&  & 300 &\(-0.170^{+0.015}_{-0.015}\) &\(0.01^{+0.07}_{-0.08}\)&\(-0.162\pm0.020\)&\(-0.77\pm0.17\)\\
\((1,0)\) & \(\Xi_{cc}K\rightarrow\Xi_{cc}K\) & 210 & \(0.715^{+0.290}_{-0.150}\) & \(0.81^{+0.21}_{-0.28}\)& \(0.697\pm0.090\) & \(1.19\pm0.20\)\\
  & & 300 & \(0.768^{+0.361}_{-0.175}\) &\(0.88^{+0.20}_{-0.26}\)&\(0.630\pm0.100\)& \(1.09\pm0.19\) \\
\((1,1)\) & \(\Xi_{cc}K\rightarrow\Xi_{cc}K\) & 210 & \(-0.171^{+0.014}_{-0.014}\) &\(0.09^{+0.08}_{-0.07}\)& \(-0.212\pm0.014\)& \(-1.44\pm0.29\) \\
 &  & 300 & \(-0.173^{+0.015}_{-0.015}\)&\(0.04^{+0.07}_{-0.07}\)&\(-0.177\pm0.022\)&\(-0.88\pm0.16\) \\
\((0,\frac{3}{2})\)& \(\Xi_{cc}\pi\rightarrow\Xi_{cc}\pi\) & 210 & \(-0.130^{+0.006}_{-0.006}\) & \(3.68^{+0.49}_{-0.41}\)&\(-0.143\pm0.024\)&\(-0.03\pm0.45\) \\
  &  & 300 &\(-0.155^{+0.009}_{-0.009}\)  & \(1.00^{+0.24}_{-0.19}\) &\(-0.140\pm0.014\)&\(-0.03\pm0.19\)\\
\hline  
\end{tabular}
\caption{Scattering lengths $a_0$ and effective ranges $r_0$ (in units of fm) for the four single-channel scattering processes at unphysical pion masses. Predictions from the UChPT and the lattice results~\cite{Yi:2025bnh} are shown together for comparisons. }\label{tab.a0r0lat} 
\end{table} 

In Ref.~\cite{Yi:2025bnh}, the quantity of $k\cot\delta(s)$ is extracted via the L\"uscher formula by using the finite-volume energy levels, where $k$ is the three-momentum of the DCB and pNGB scattering in the CM frame and $\delta$ represents the phase shift.
By further exploiting the effective range expansion (ERE), the scattering parameters at threshold, viz. the $S$-wave scattering length ($a_0$) and effective range ($r_0$), are determined as well. The ERE convention adopted in the lattice study reads 
\begin{equation}\label{eq.yy2}
\begin{aligned}
    k\,\text{cot}\delta(s)=\frac{1}{a_{0}}+\frac{1}{2}r_{0}k^2+O(k^4)\,,
\end{aligned}
\end{equation}
where $k=\sqrt{2\mu (\sqrt{s}-m_{\text{th}})}$ stands for the non-relativistic (NR) CM three momentum, with $m_{\mathrm{th}}=m_1+m_2$ and $\mu=m_1 m_2/(m_{1}+m_{2})$. The $S$-wave scattering lengths and effective ranges calculated from the lattice study are listed in Table \ref{tab.a0r0lat}. 

The ERE parameters can also be extracted from the chiral unitarized amplitudes by performing the NR momentum expansion near threshold 
\begin{equation}\label{eq.kcotduchpt}
\begin{aligned}
8\pi\sqrt{s}\bigg[\frac{1}{T^{\text{UChPT}}(s)}+i\rho(s)\bigg]=\frac{1}{a_{0}}+\frac{1}{2}r_{0}k^2 +O(k^4)\,,
\end{aligned}
\end{equation}
where $8\pi\sqrt{s}$ is introduced due to the different normalization between the relativistic and NR amplitudes. The left-hand side of Eq.~\eqref{eq.kcotduchpt} corresponds to the quantity of $k\cot\delta(s)$ calculated in the unitarized chiral approach, which will be further used to fit the lattice data provided by Ref.~\cite{Yi:2025bnh}.  

\begin{figure}[htpb]
  \centering
  \begin{subfigure}[b]{\textwidth}
    \centering
    \includegraphics[width=0.44\textwidth,height=4.2cm]{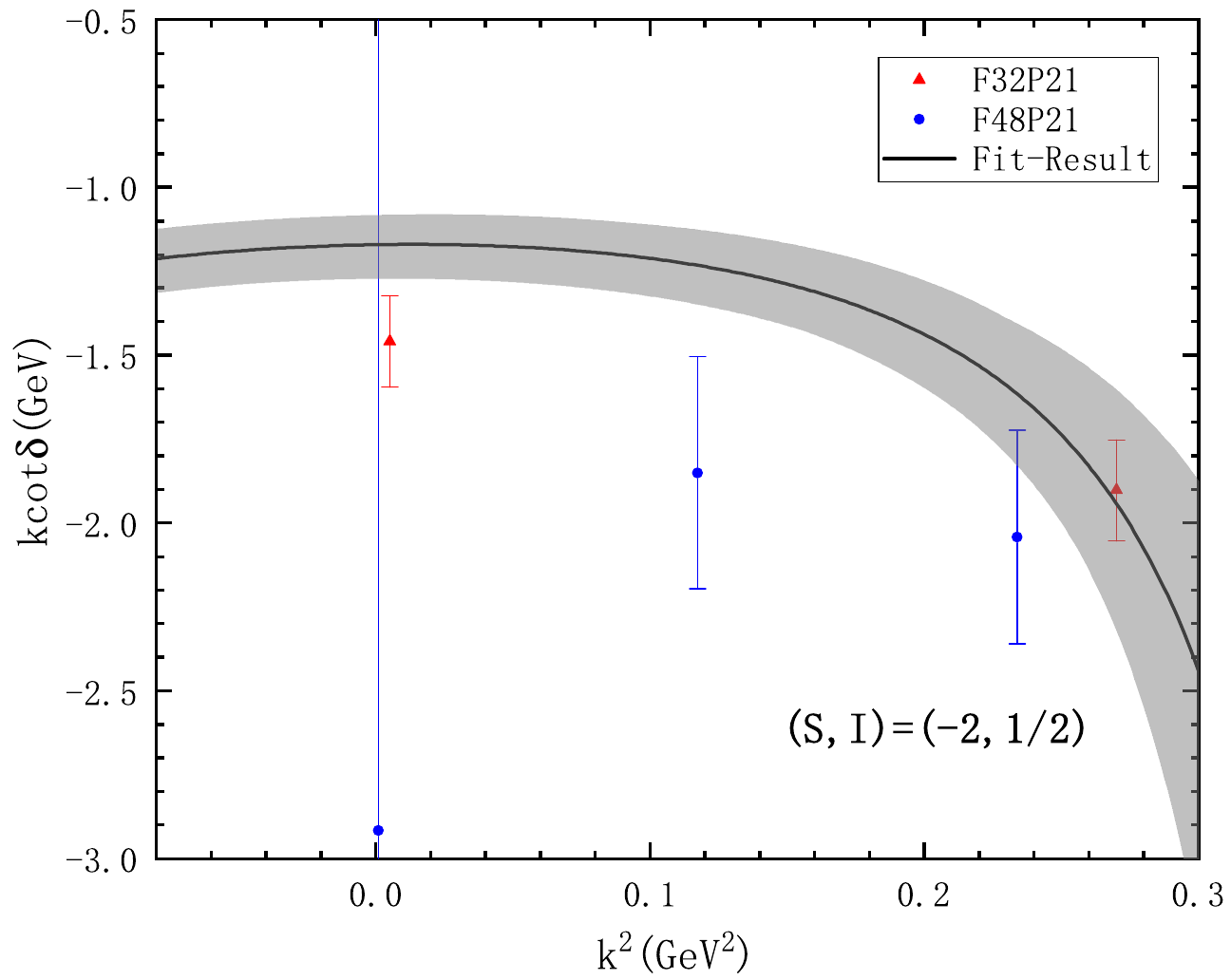}\hfill
        \includegraphics[width=0.44\textwidth,height=4.2cm]{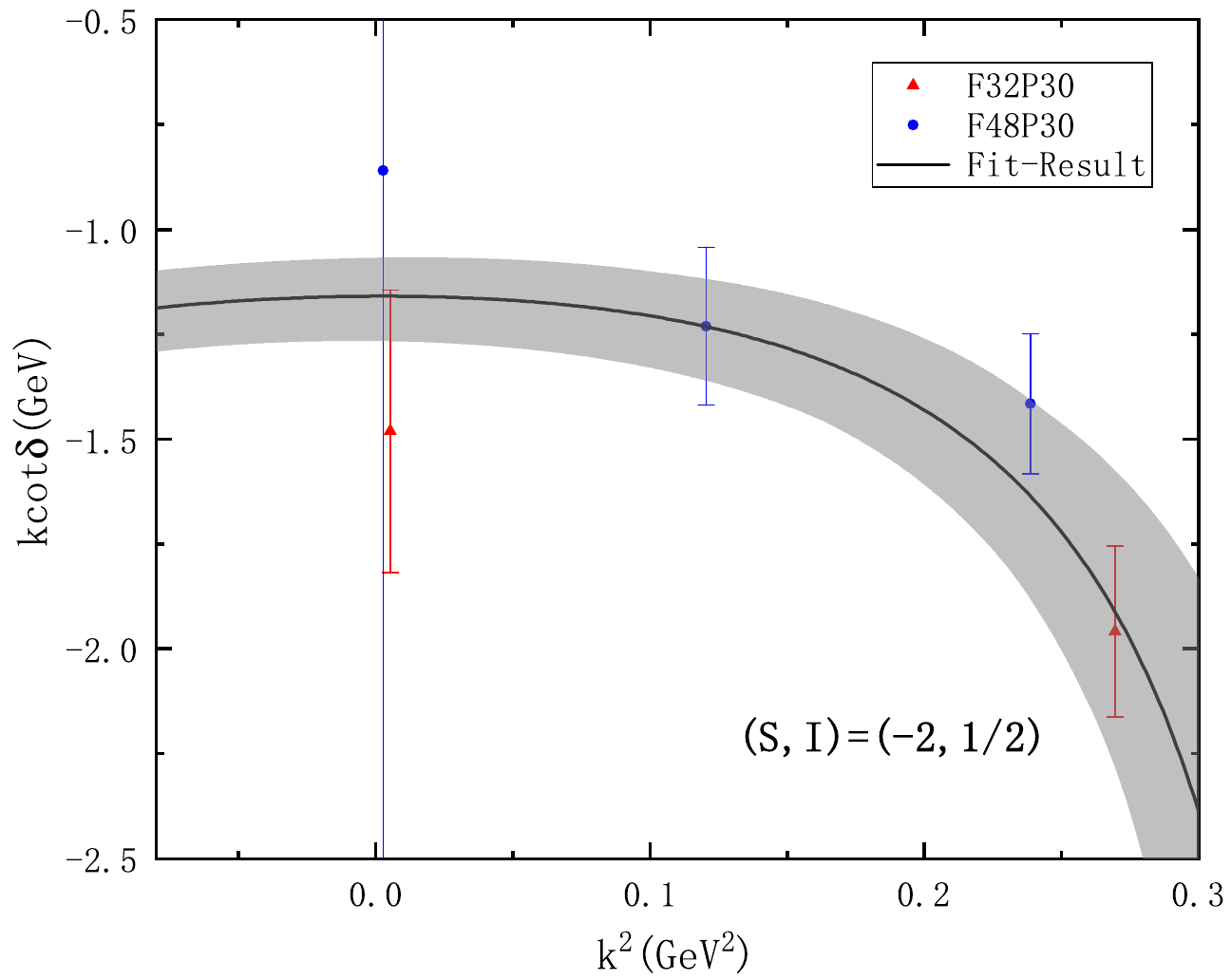}
  \end{subfigure}  
  \vskip\baselineskip
  \begin{subfigure}[b]{\textwidth}
    \centering
    \includegraphics[width=0.44\textwidth,height=4.2cm]{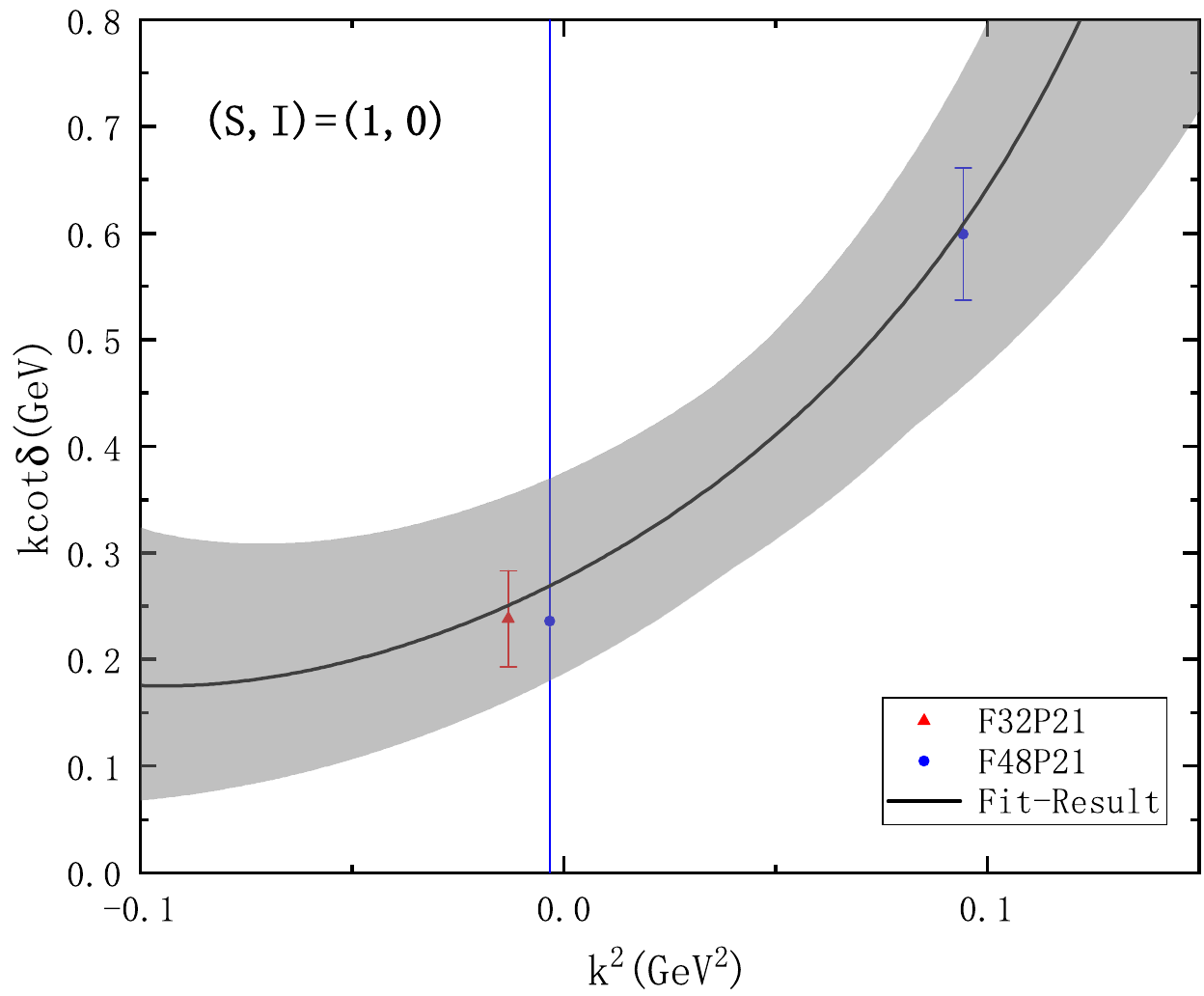}\hfill
        \includegraphics[width=0.44\textwidth,height=4.2cm]{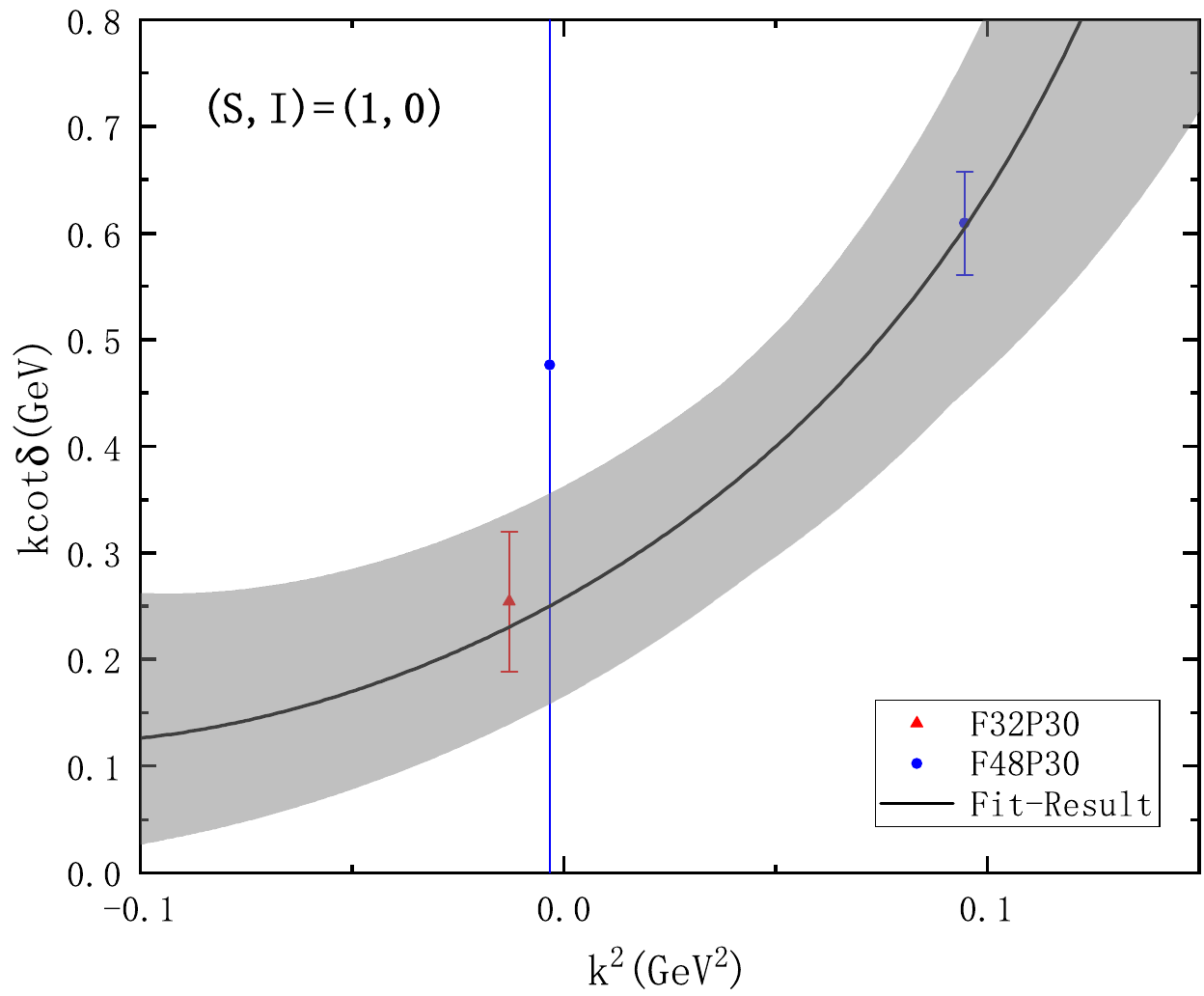}
  \end{subfigure}
  \vskip\baselineskip
  \begin{subfigure}[b]{\textwidth}
    \centering
    \includegraphics[width=0.44\textwidth,height=4.2cm]{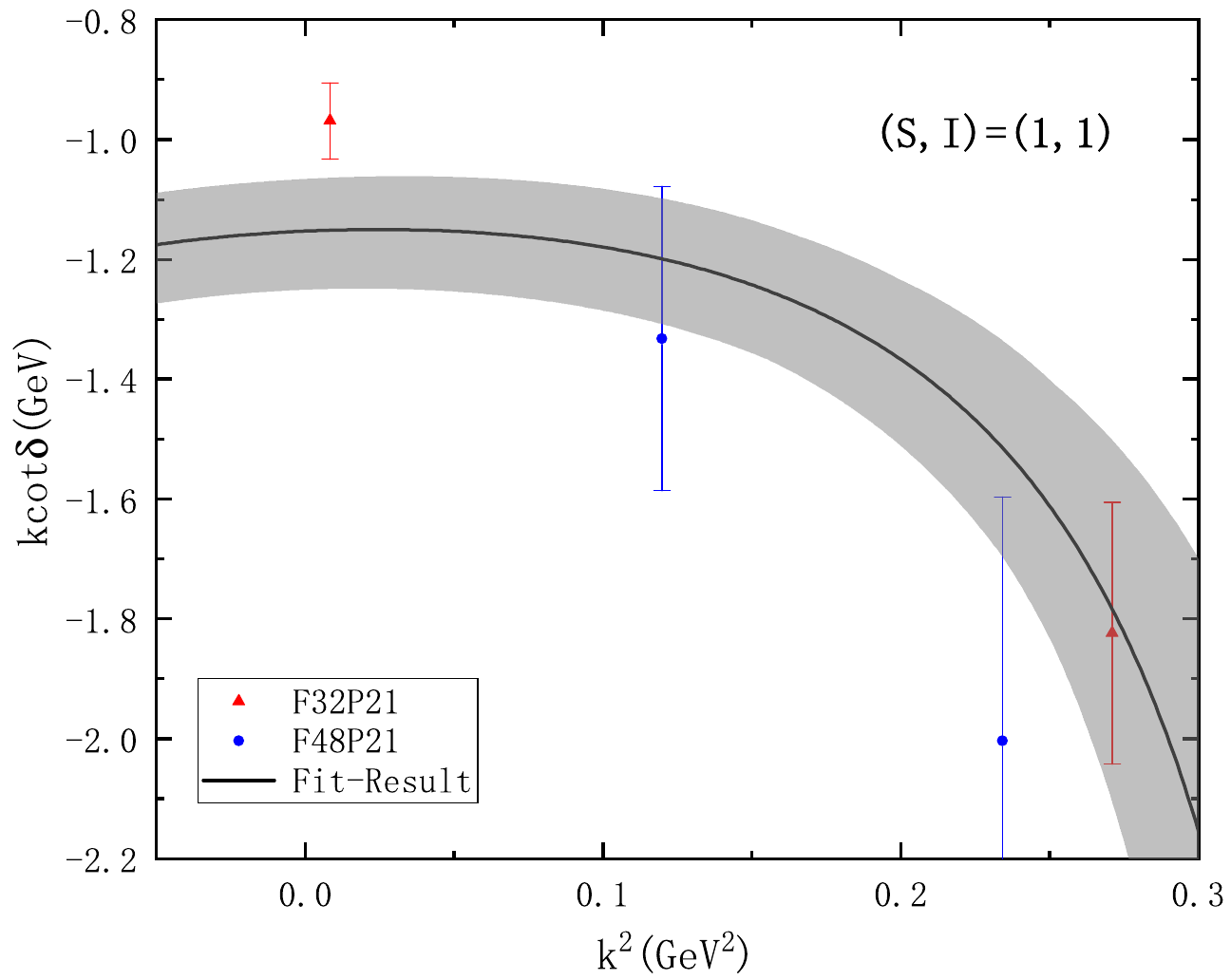}\hfill
        \includegraphics[width=0.44\textwidth,height=4.2cm]{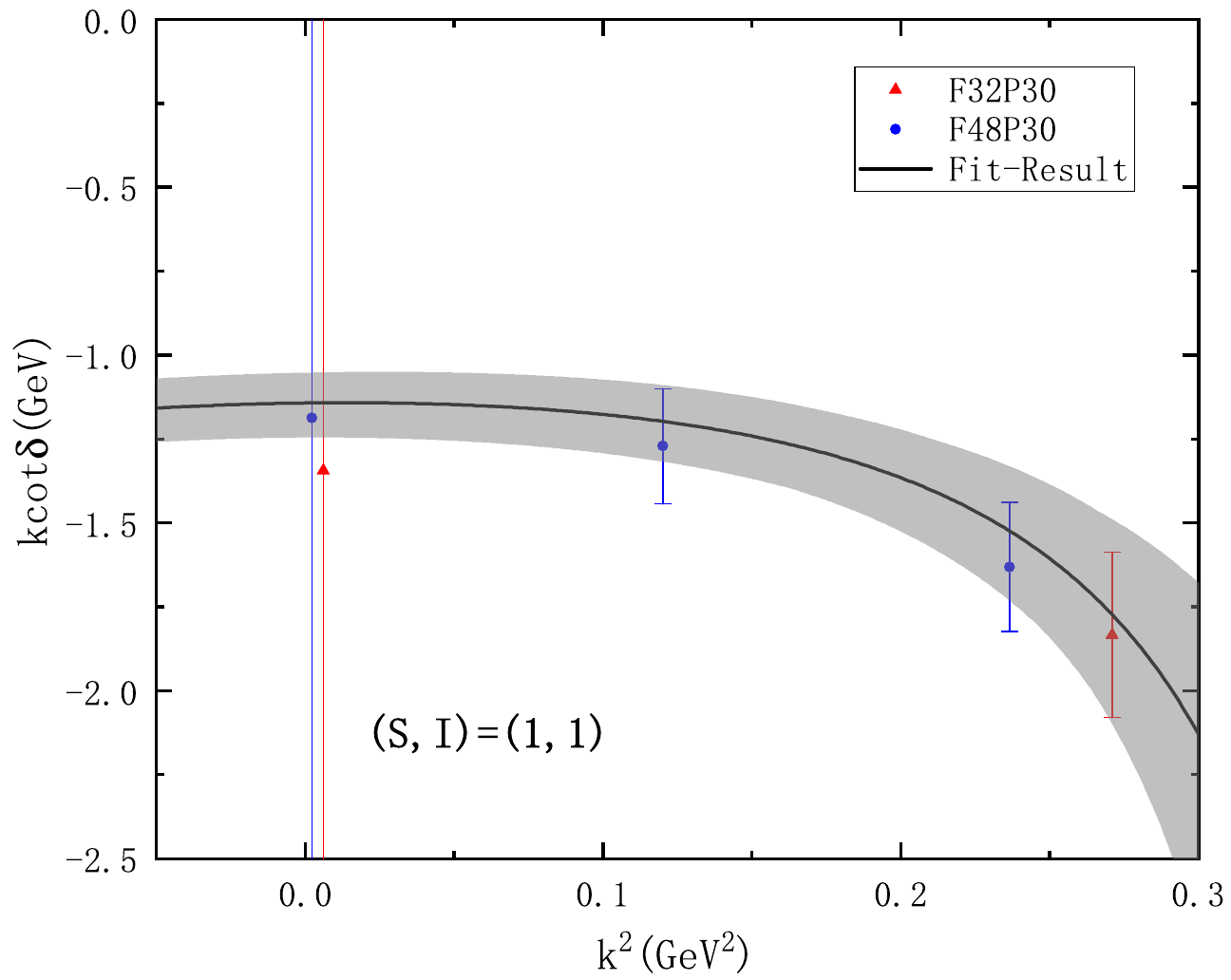}
  \end{subfigure}  
  \vskip\baselineskip
  \begin{subfigure}[b]{\textwidth}
    \centering
    \includegraphics[width=0.44\textwidth,height=4.2cm]{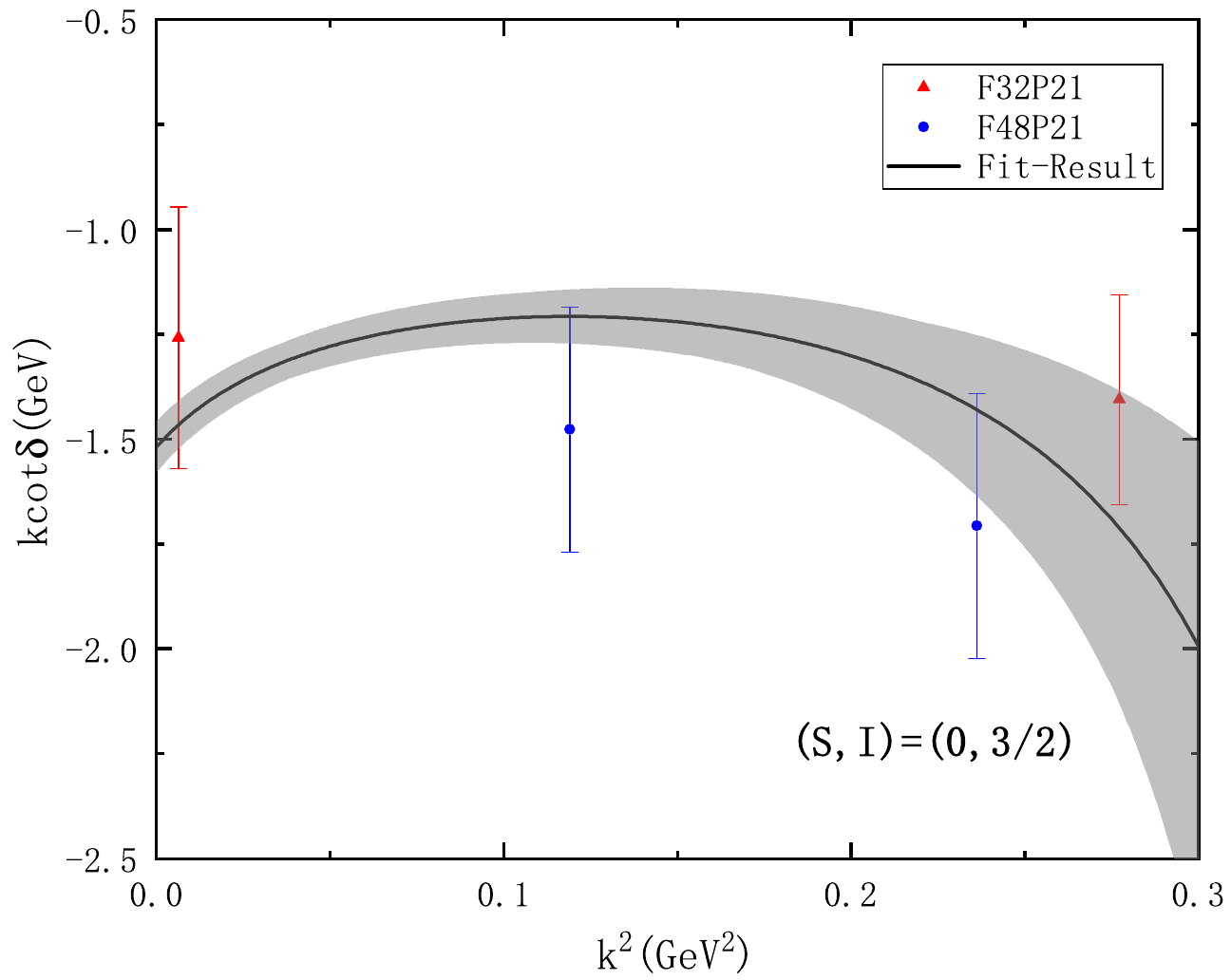}\hfill
        \includegraphics[width=0.44\textwidth,height=4.2cm]{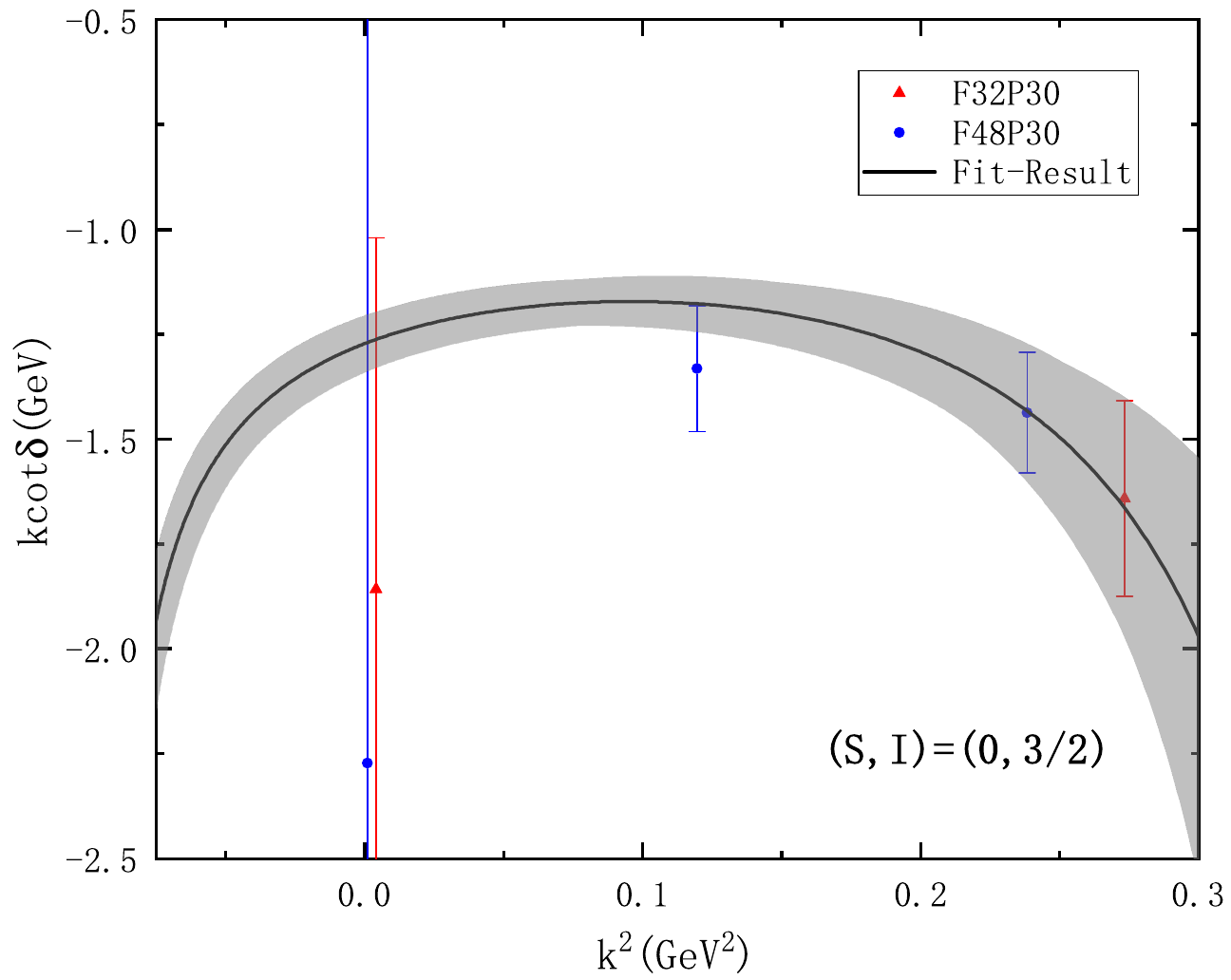}
  \end{subfigure}
  \caption{Fitted results of $k\,\text{cot}\delta$ for the $(S,I)=(-2,\frac{1}{2})$, $(1,0)$, $(1,1)$ and $(0,\frac{3}{2})$ channels at pion masses $m_{\pi}\sim 210\,\text{MeV}$ and $300\,\text{MeV}$. The black solid curves show the best-fit results, and the gray shaded bands indicate the corresponding uncertainties. The data points are obtained from lattice QCD calculations~\cite{Yi:2025bnh}.}\label{fig.fitpcotd} 
  \end{figure}

The axial coupling constant is fixed to $g=-0.19$~\cite{Liang:2023scp}, while the remaining parameters, including the cutoff $\Lambda$ and the seven NLO LECs $b_{i=1,\cdots,7}$, are taken as free parameters in the fits. The values of various parameters from our best fit are 
\begin{equation}\label{eq.fitlecs}
\begin{aligned}
    b_{1}&=0.04\pm0.01\,\text{GeV}^{-1}\,,\quad & b_{2}&=-0.01\pm0.03\,\text{GeV}^{-1}\,,\\
     b_{3}&=-1.28\pm0.02\,\text{GeV}^{-1}\,,\quad & b_{4}&=0.55\pm0.01\,\text{GeV}^{-1}\,,\\
      b_{5}&=-0.16\pm0.01\,\text{GeV}^{-1}\,,\quad & b_{6}&=0.13\pm0.01\,\text{GeV}^{-1}\,,\\
       b_{7}&=0.78\pm0.08\,\text{GeV}^{-1}\,,\quad & \Lambda&=762.9\pm 57.6\,\text{MeV} \,,
\end{aligned}
\end{equation}
with $\chi^2/{\text{d.o.f}}=\frac{28.1}{36-8}\simeq 1.0$. It is noted that a universal value of $\Lambda$ is taken for the involved scattering processes. We have tried to free in the fit the axial coupling $g$, which is found to be unstable, and the fit quality after releasing the parameter $g$ is barely changed, compared to the case by fixing its value at $-0.19$ as adopted in Ref.~\cite{Liang:2023scp}. As a result, we consider the fit result by fixing $g=-0.19$ as our preferred one in this work.

The uncertainties of the parameters are obtained by employing the bootstrap method. Namely, huge amount of pseudo lattice data are randomly generated by assuming the Gaussian distributions according to their means and standard deviations from the lattice study in Ref.~\cite{Yi:2025bnh}. We then redo the fits by taking each set of the pseudo lattice data and keep the parameters from the fits at the 1-$\sigma$ confidence level. 
The remaining uncertainty analyses, including the error bands of various plots, scattering parameters and pole positions, are based on the parameter sets from this bootstrap procedure.  
The values of $b_{i=1,\cdots,6}$ from our study are compatible within uncertainties with the predictions based on HADS~\cite{Liang:2023scp} and the determinations by fitting the masses of DCBs on the lattice~\cite{Yao:2018ifh}, indicating that HADS works well in the DCB system. The value of $b_7$ is simply set as zero in Ref.~\cite{Liang:2023scp}, since it can not be fixed by the HADS. The fit in our study clearly gives a non-vanishing value for $b_7$.

The resulting curves of the $k\,\text{cot}\delta(s)$ at different pion masses are shown in Fig.~\ref{fig.fitpcotd}, together with the lattice data from Ref.~\cite{Yi:2025bnh}. It is noted that the lattice simulations are performed by taking the ensembles with almost two degenerate pion masses at two different volumes. To be specific, we have plotted the theoretical curves with $m_\pi= 303.96\,\mathrm{MeV}$ and $208.50\,\mathrm{MeV}$. The results by taking another almost degenerate masses with $m_\pi= 304.87\,\mathrm{MeV}$ and $207.74\,\mathrm{MeV}$ are indistinguishable and hence are not explicitly shown. The masses of $K$, $\Xi_{cc}$ and $\Omega_{cc}$ employed in the lattice simulations in Ref.~\cite{Yi:2025bnh} can be seen in Table~\ref{tab.latensem}.

According to the plots in Fig.\ref{fig.fitpcotd}, the fits can well reproduce almost all the lattice data points. The scattering lengths $a_0$ and effective ranges $r_0$ calculated from the unitarized chiral amplitudes via Eq.\eqref{eq.kcotduchpt} are listed in Table \ref{tab.a0r0lat}, where we also show the determinations from the lattice study in the last two columns for comparison. Overall, after taking into account the uncertainties, the values of $a_0$ from our study agree with those from the lattice analyses based on the LO ERE formalism. However, regarding the effective ranges $r_0$, clear discrepancies are observed between our determinations and the LO ERE results from the lattice study. In fact, the lattice study finds that only the scattering lengths $a_0$ can be robustly determined, while the effective ranges $r_0$ can not be accurately fixed. E.g., the values of $r_0$ are only explicitly quoted for the LO ERE analyses in Ref.~\cite{Yi:2025bnh} and it is stated that the errors of $r_0$ from the NLO ERE are too large to be shown. Our predictions to the effective ranges $r_0$ provide useful references for future more precise lattice calculations.

\subsection{Extrapolation of scattering amplitudes to physical quark masses}

Through the fits to lattice data at unphysically large quark masses, the values of the unknown LECs are determined in Eq.~\eqref{eq.fitlecs}. Since those LECs are independent of the quark masses, this allows us to perform the chiral extrapolation to physical masses. 
Therefore, in this part we give the predictions to the quantities related to the scattering amplitudes at physical masses, such as  the scattering lengths, effective ranges, phase shifts and inelasticities.

The resulting scattering lengths and effective ranges at physical masses by unitarizing the NLO chiral amplitudes are presented in Table~\ref{tab.phya0r0}, together with the results from the LO unitarized chiral amplitudes~\cite{Guo:2017vcf} and the NLO perturbative chiral amplitudes~\cite{Liang:2023scp}. 
For the elastic channels with $(S,I)=(-2,\frac{1}{2})$, $(1,1)$ and $(0,\frac{3}{2})$, the values of $a_0$ from the present study agree with the LO unitarized results in Ref.~\cite{Guo:2017vcf} within uncertainties. 
For the $\Xi_{cc}K$ channel with $(S,I)=(1,0)$, the unitarized results by taking the LO and NLO amplitudes as inputs give rather different values for $a_0$. 
In contrast to the LO unitarized study in Ref.~\cite{Guo:2017vcf}, where the cutoff $\Lambda$ is simply chosen in a wide range as a guess, the parameters entering the unitarized NLO amplitudes in the present work are determined by the fits to the lattice data, which should be considered more reliable. Furthermore, the LO unitarized result is highly sensitive to the three-momentum cutoff $\Lambda$. The scattering length can become rather large due to a pole approaching the $\Xi_{cc}K$ threshold for $\Lambda\approx 900\,\mathrm{MeV}$. After including NLO corrections, the scattering length stabilizes, and the result at the physical pion mass is consistent with that at the lattice pion mass.
Therefore the results in this study should supersede the ones from the LO unitarized calculations in Ref.~\cite{Guo:2017vcf}.  
Alternatively, Ref.~\cite{Liang:2023scp} has computed the scattering lengths from the perturbative NLO chiral expressions by using the LECs estimated from HADS. The results are also listed in the last column of Table~\ref{tab.phya0r0} for comparison. In addition to the single-channel case, we also predict the values of $a_0$ and $r_0$ for the coupled-channel scattering with $(S,I)=(-1,0),(-1,1)$ and $(0,1/2)$.

\begin{figure}[htpb]
  \centering
  \begin{subfigure}[b]{0.48\textwidth}
    \centering
    \includegraphics[width=\textwidth]{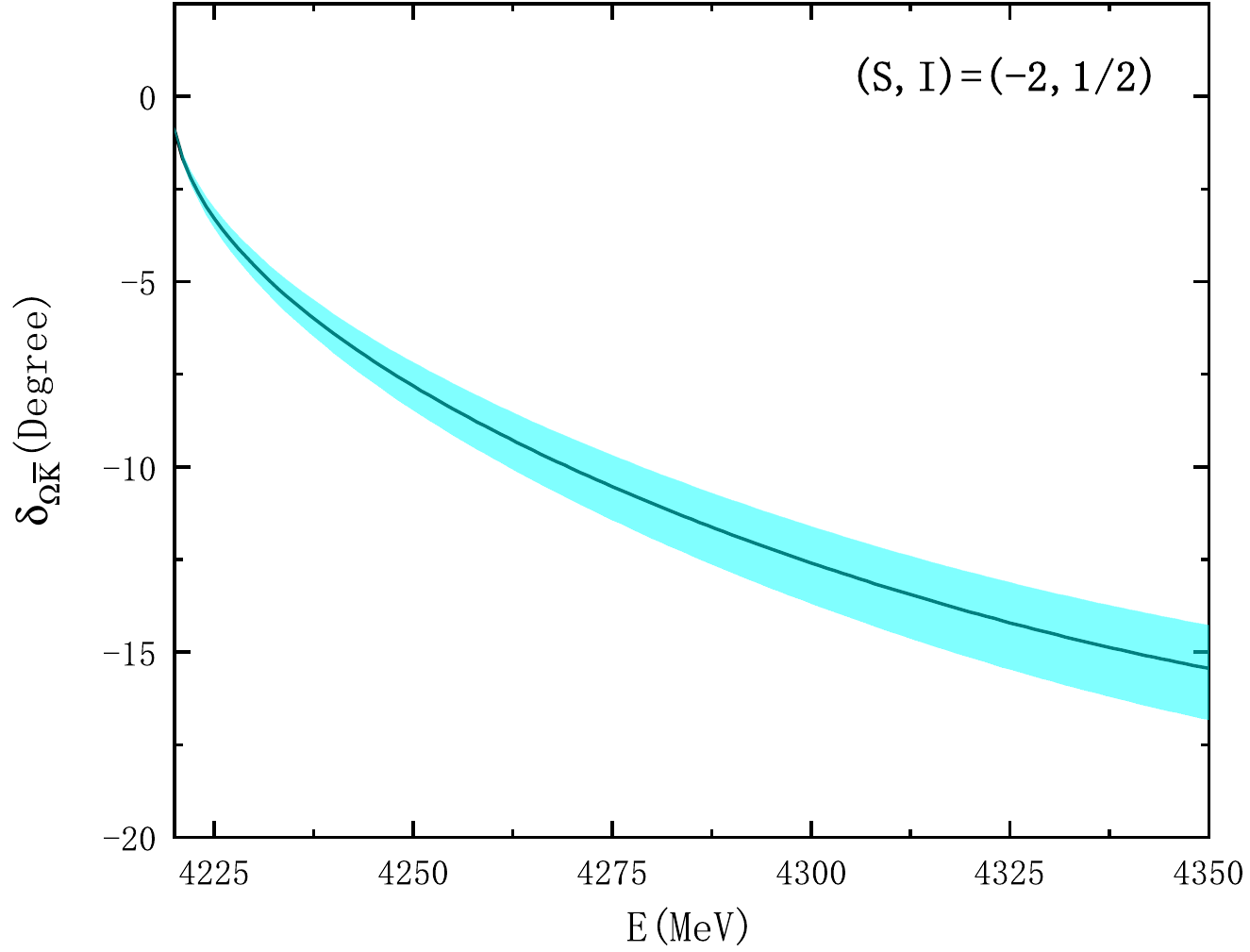}
    \captionsetup{justification=centering}
  \end{subfigure}
  \hfill
  \begin{subfigure}[b]{0.48\textwidth}
    \centering
    \includegraphics[width=\textwidth]{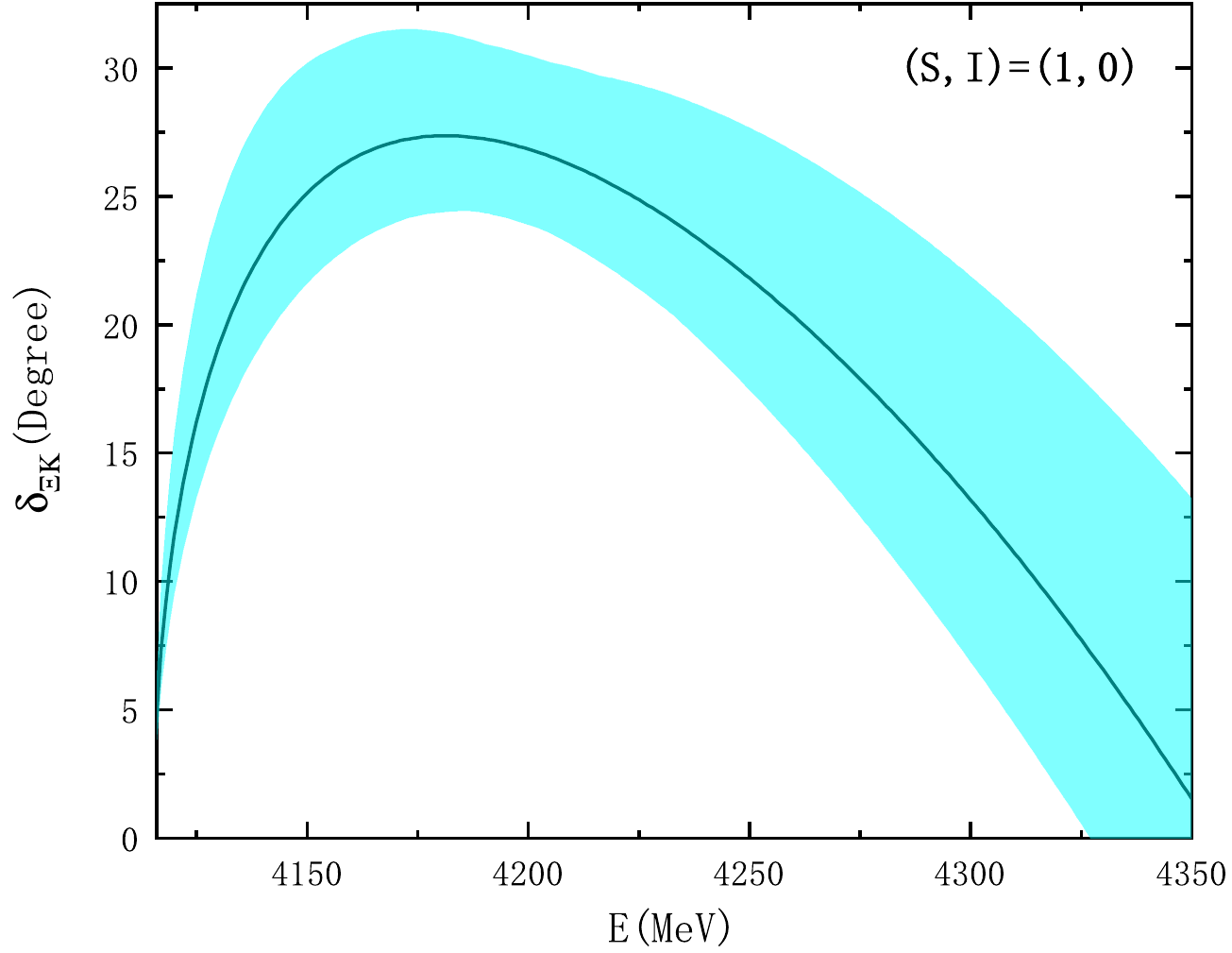}
    \captionsetup{justification=centering}
  \end{subfigure}
  \vskip\baselineskip
  \begin{subfigure}[b]{0.48\textwidth}
    \centering
    \includegraphics[width=\textwidth]{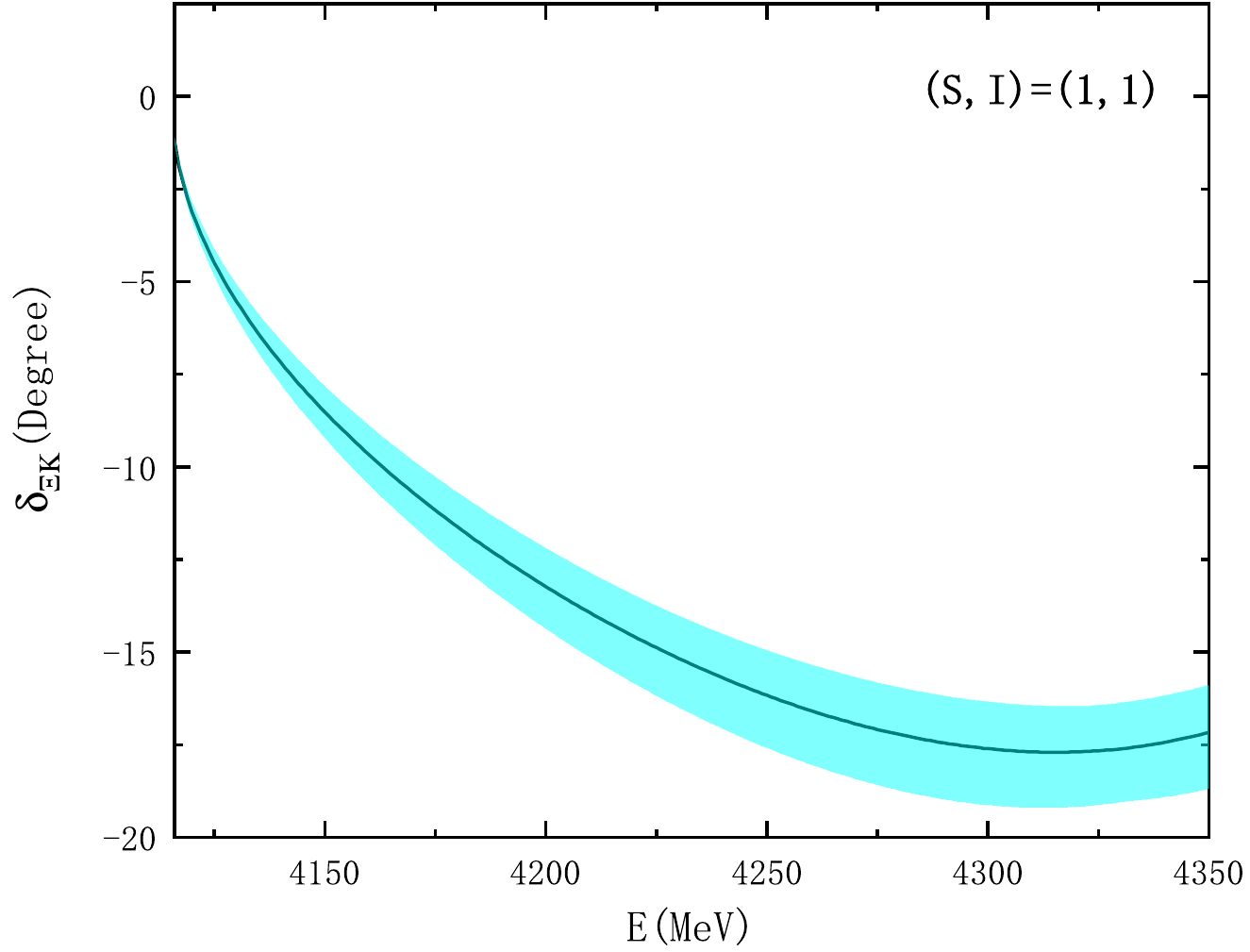}
    \captionsetup{justification=centering}
  \end{subfigure}
  \hfill
  \begin{subfigure}[b]{0.48\textwidth}
    \centering
    \includegraphics[width=\textwidth]{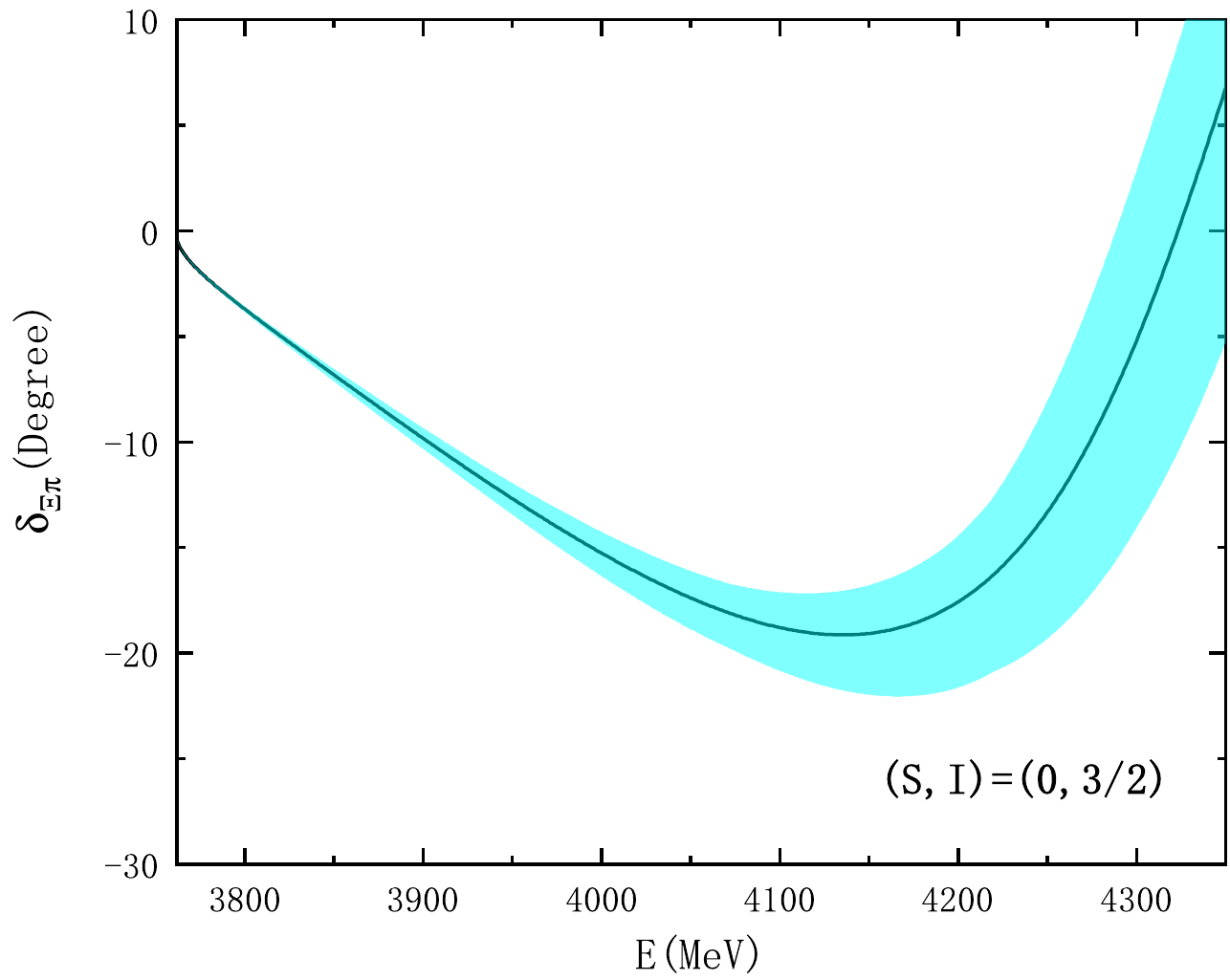}
    \captionsetup{justification=centering}
  \end{subfigure}
  \caption{$S$-wave phase shifts for the elastic scattering cases. The black solid curves represent the phase shifts from the best fit in Eq.~\eqref{eq.fitlecs}, and the blue shaded bands indicate the uncertainties.}\label{fig.scphaseshifts}
\end{figure}

\begin{figure}[htbp]
  \centering
  \begin{subfigure}[b]{0.48\textwidth}
    \centering
    \includegraphics[width=\textwidth]{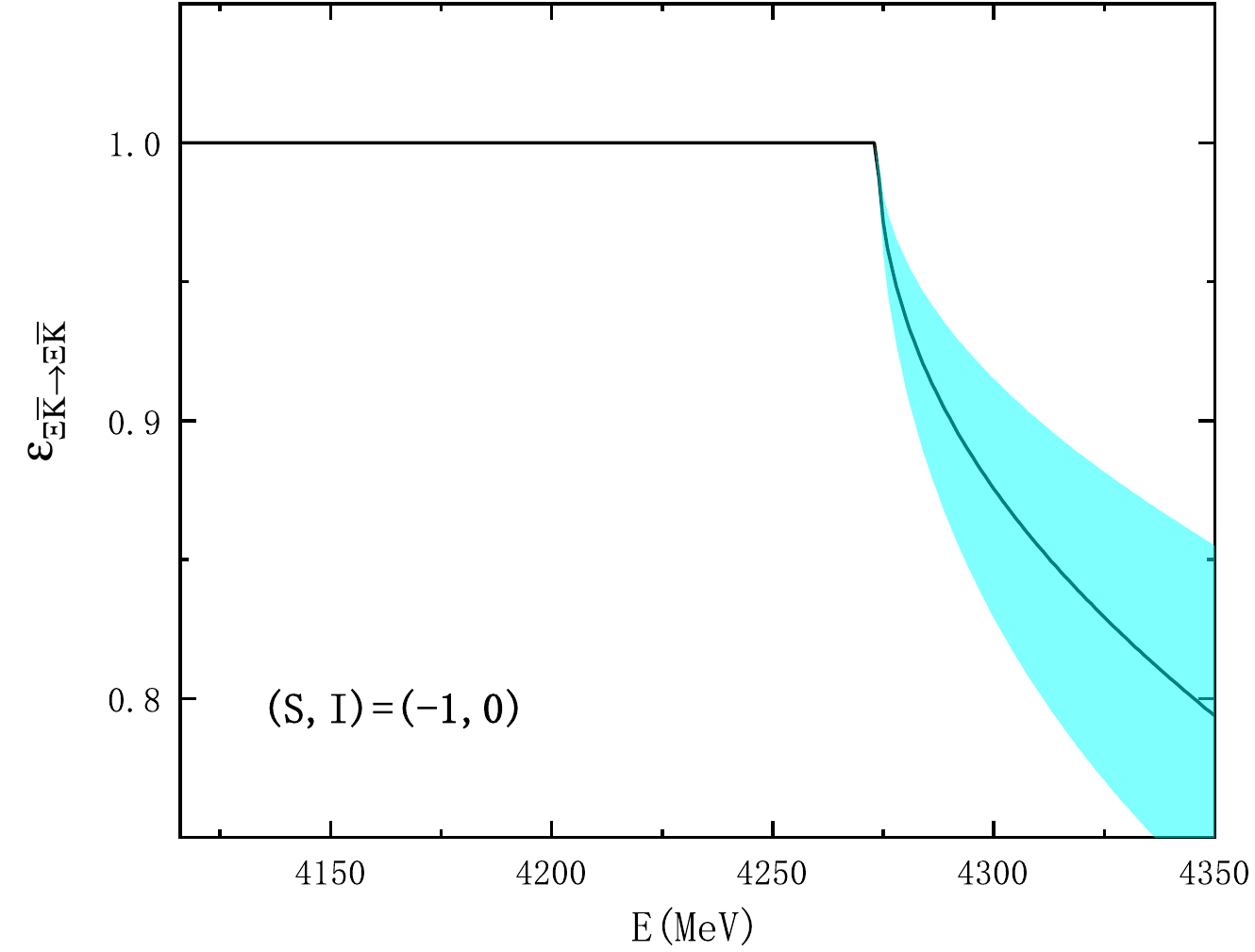}
    \captionsetup{justification=centering}
  \end{subfigure}
  \hfill
  \begin{subfigure}[b]{0.48\textwidth}
    \centering
    \includegraphics[width=\textwidth]{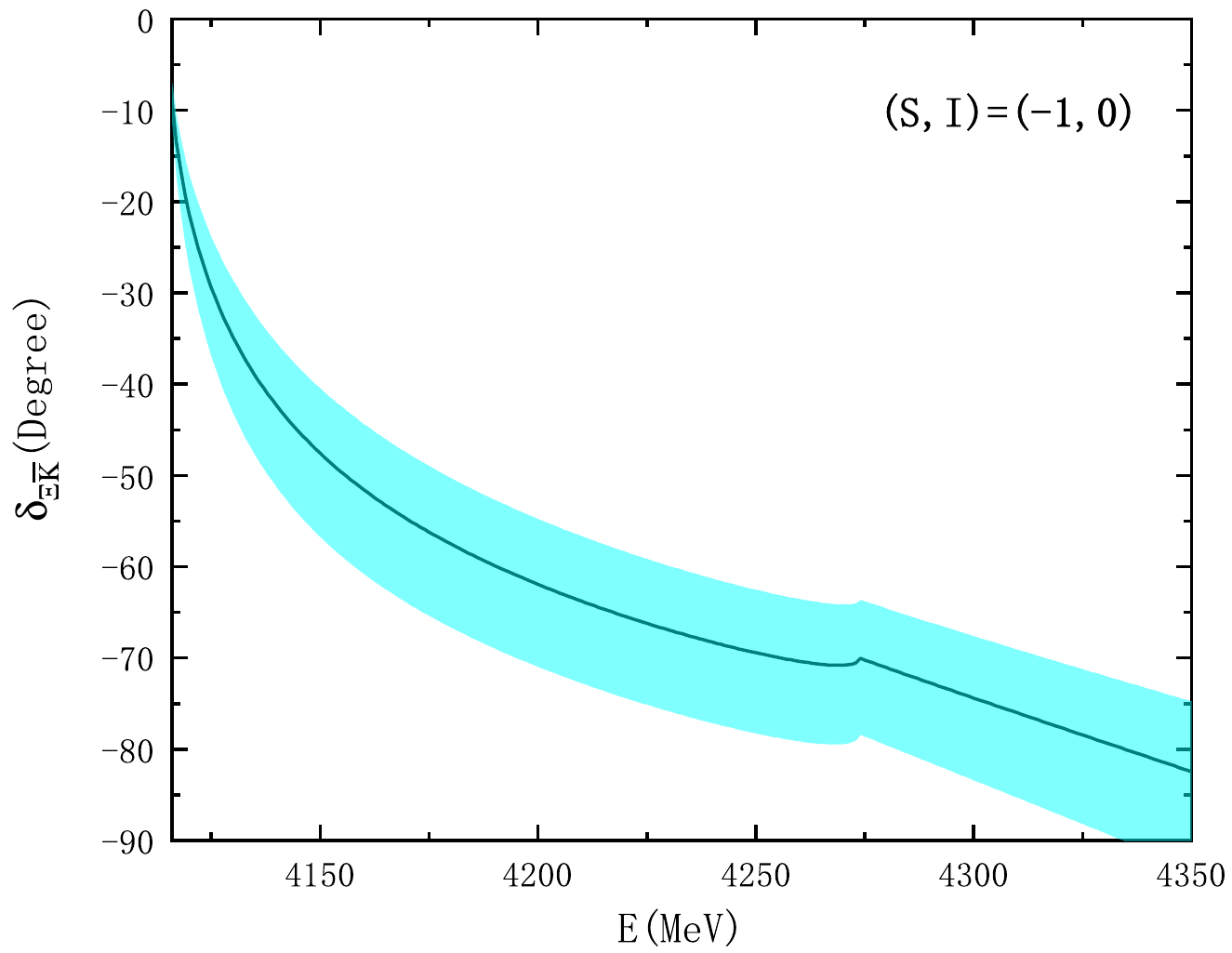}
    \captionsetup{justification=centering}
  \end{subfigure}
  \vskip\baselineskip
  \begin{subfigure}[b]{0.48\textwidth}
    \centering
    \includegraphics[width=\textwidth]{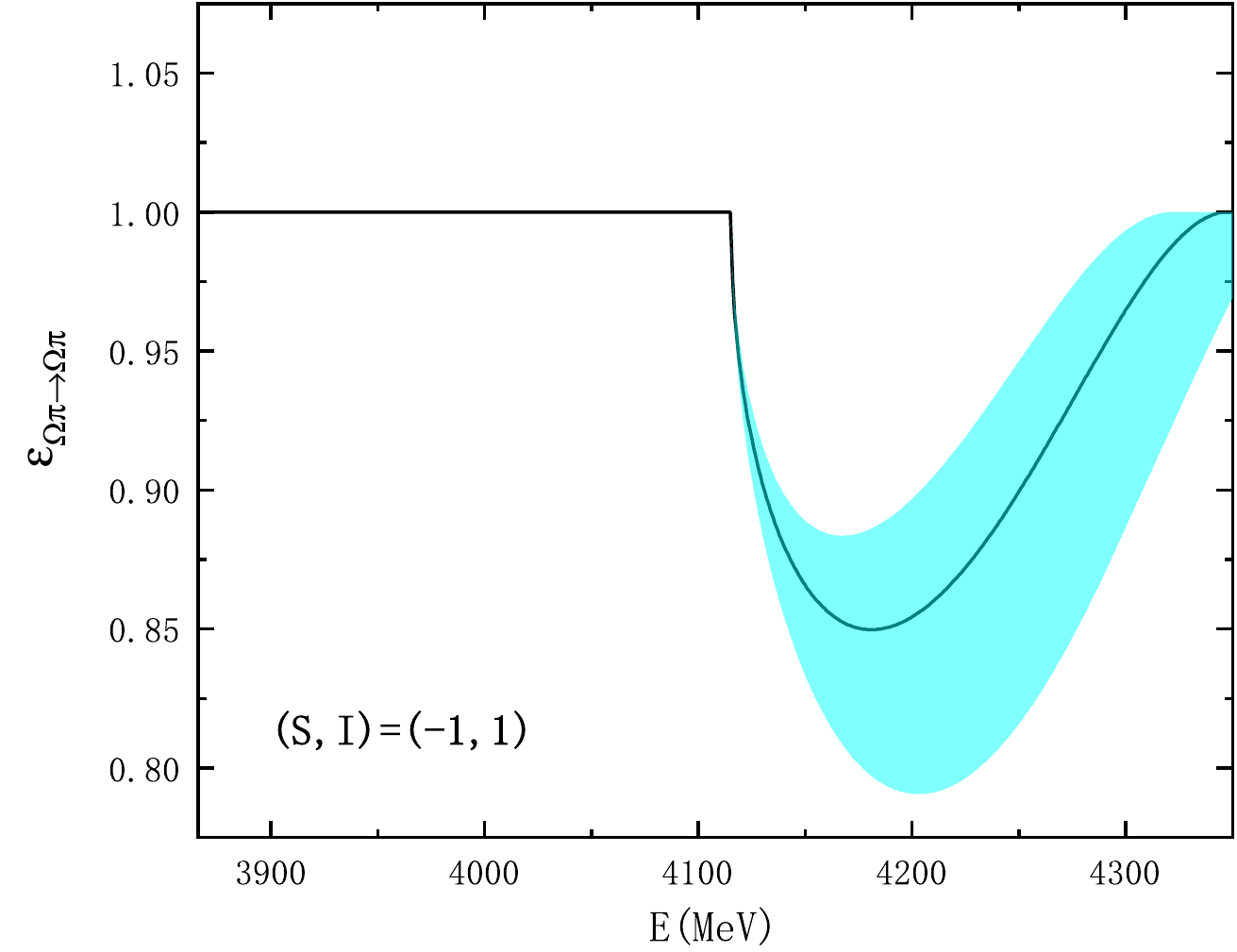}
    \captionsetup{justification=centering}
  \end{subfigure}
  \hfill
  \begin{subfigure}[b]{0.48\textwidth}
    \centering
    \includegraphics[width=\textwidth]{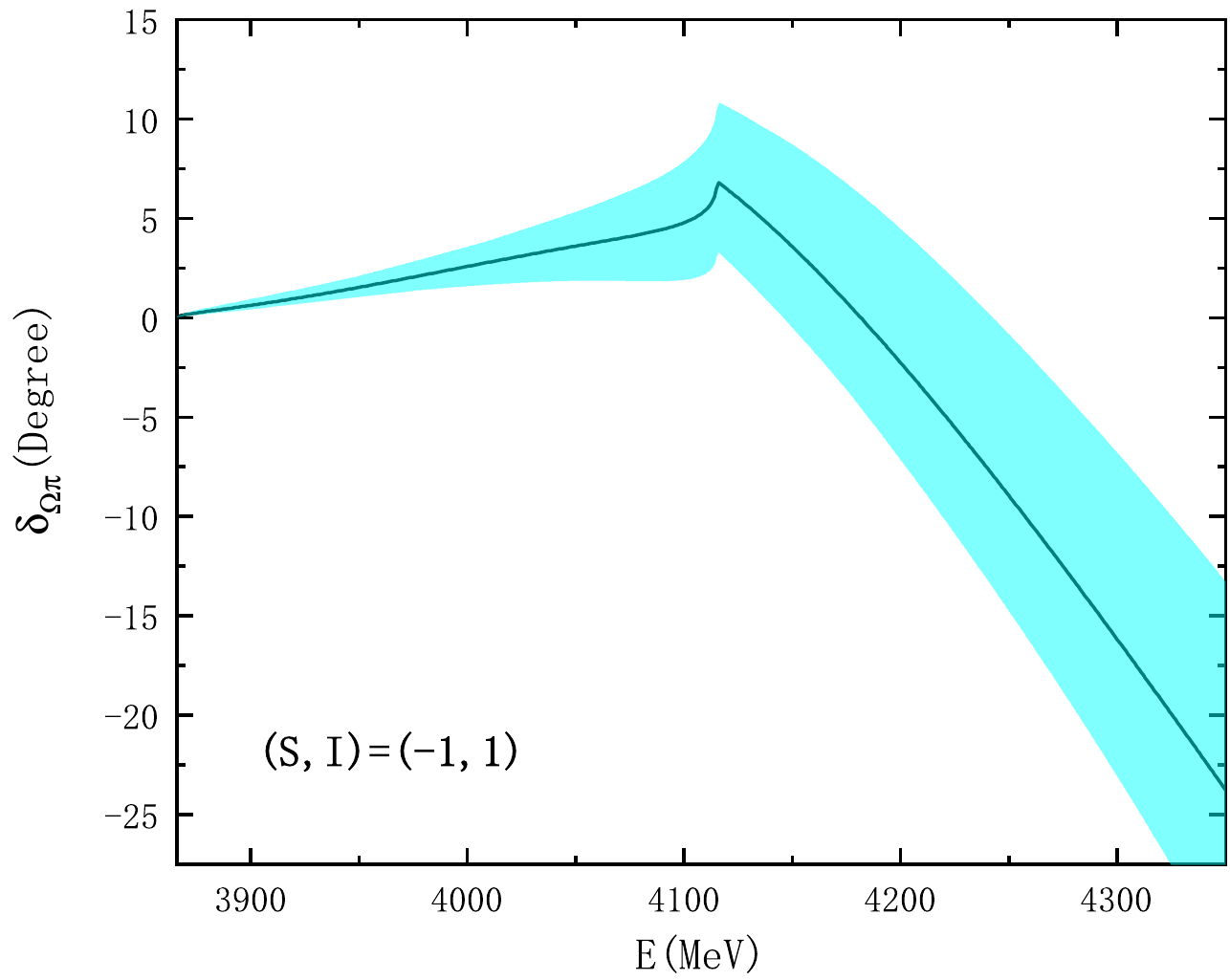}
    \captionsetup{justification=centering}
  \end{subfigure}
   \vskip\baselineskip
   \begin{subfigure}[b]{0.48\textwidth}
    \centering
    \includegraphics[width=\textwidth]{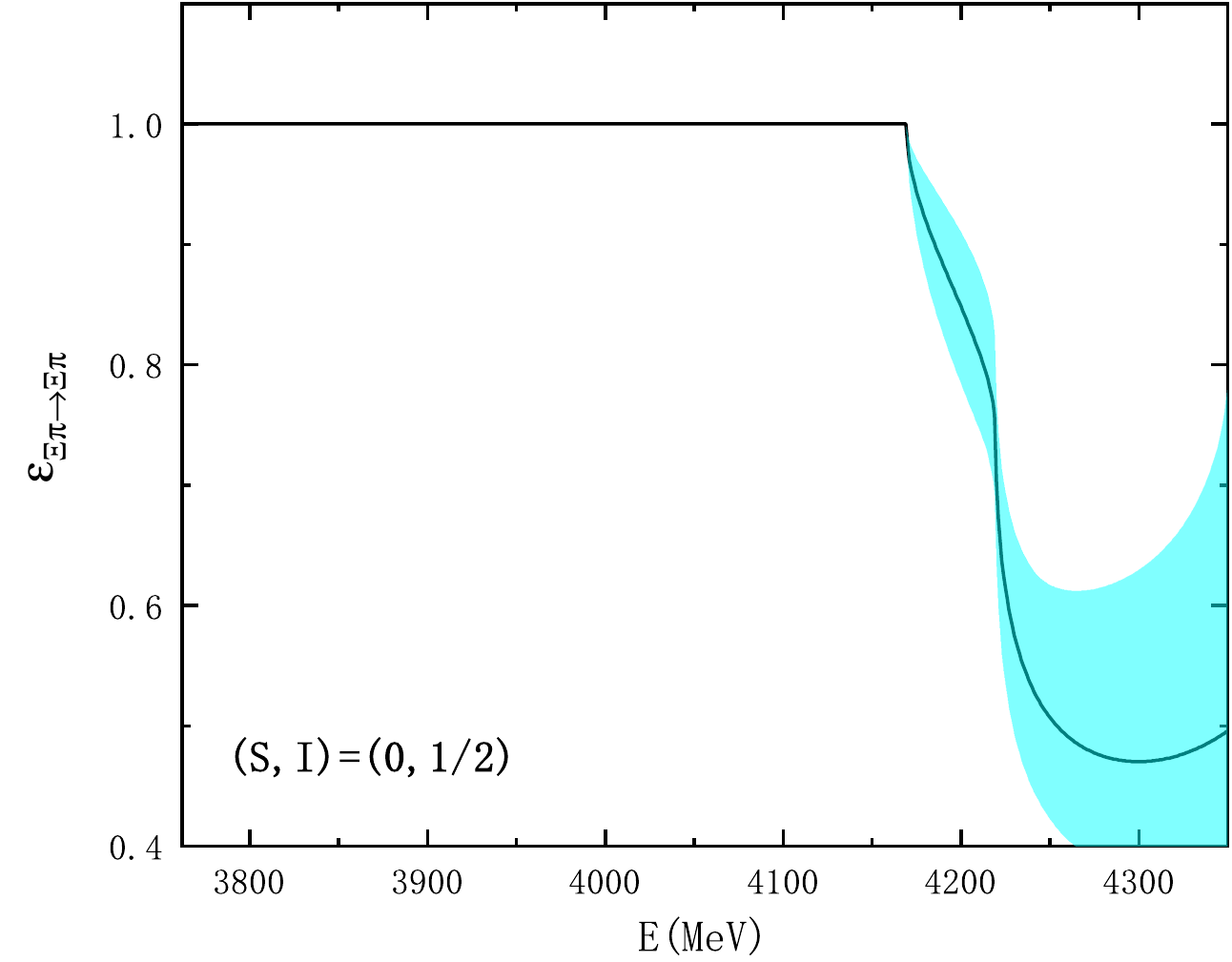}
    \captionsetup{justification=centering}
  \end{subfigure}
  \hfill
  \begin{subfigure}[b]{0.48\textwidth}
    \centering
    \includegraphics[width=\textwidth]{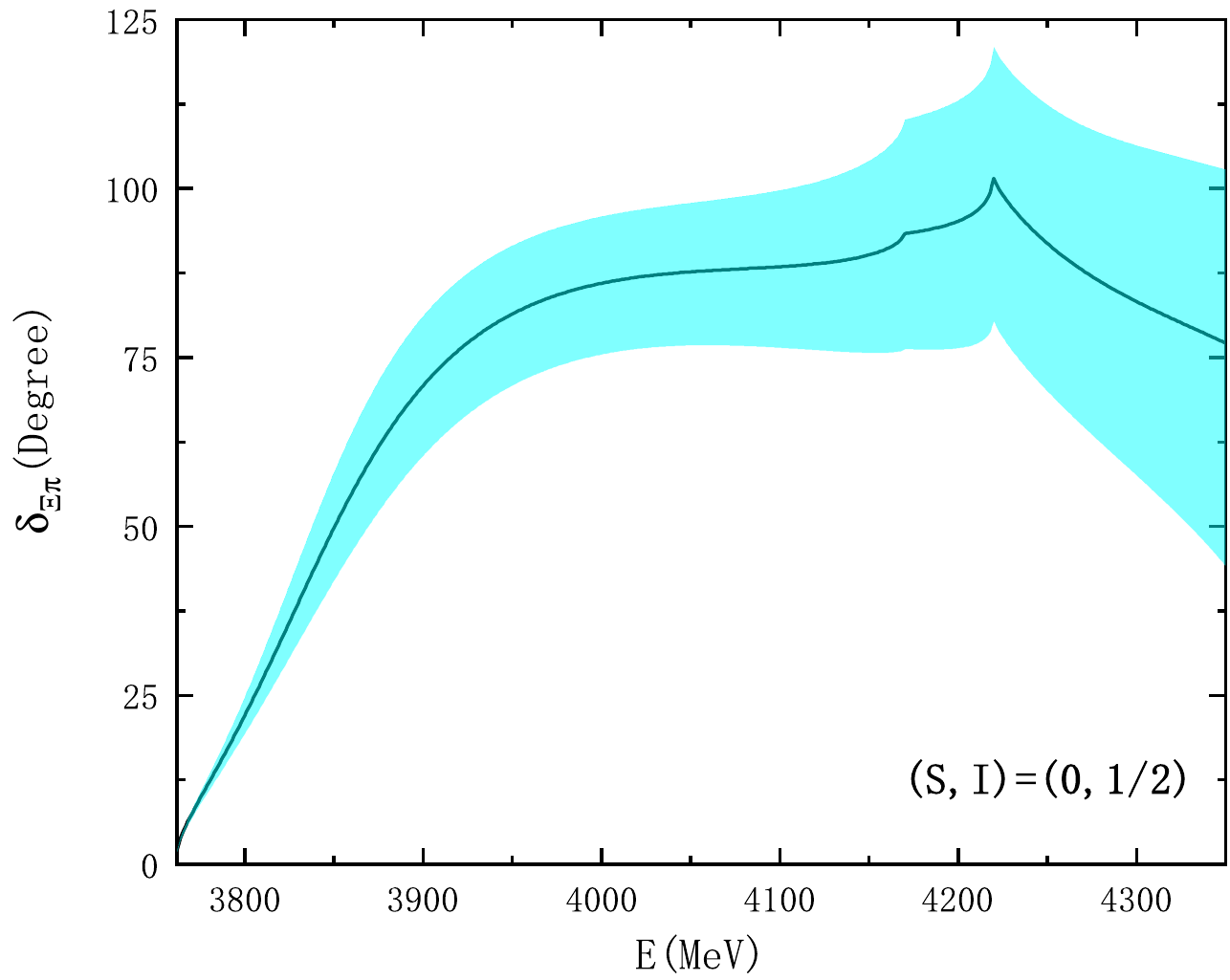}
    \captionsetup{justification=centering}
  \end{subfigure}
  \caption{$S$-wave inelasticity parameters and phase shifts for the coupled-channel scattering cases. The black solid curves represent the phase shifts from the best fit, and the blue shaded bands indicate the uncertainties.  
  }\label{fig.ccphaseshifts}
\end{figure}

The phase shifts for the four single-channel scattering processes are illustrated in Fig.~\ref{fig.scphaseshifts}. The inelasticities and phase shifts for the coupled-channel processes are shown in Fig.~\ref{fig.ccphaseshifts}. It is noted that all the phase shifts and inelasticities in Figs.~\ref{fig.scphaseshifts} and \ref{fig.ccphaseshifts} correspond to the results obtained at physical masses.

\begin{table} [htbp] 
\centering
\renewcommand{\arraystretch}{1.5}
\scriptsize
\begin{tabular}{c c c c c c}  
\hline  
\((S,I)\) & Processes &  \(a_0\) & \(r_0\) & \(a_0\)(Ref.~\cite{Guo:2017vcf}) & \(a_0\)(Ref.~\cite{Liang:2023scp}) \\
\hline  
\((-2,\frac{1}{2})\) & \(\Omega_{cc}\bar{K}\rightarrow\Omega_{cc}\bar{K}\) & \(-0.165^{+0.014}_{-0.014}\) & \(0.03^{+0.08}_{-0.09}\) & \(-0.19^{+0.02}_{-0.02}\)&0.02\\
\((1,0)\) & \(\Xi_{cc}K\rightarrow\Xi_{cc}K\) & \(0.661^{+0.249}_{-0.133}\) & \(0.83^{+0.22}_{-0.29}\)&\(-3.6 \)&0.61\\  
\((1,1)\) & \(\Xi_{cc}K\rightarrow\Xi_{cc}K\) & \(-0.168^{+0.014}_{-0.014}\) & \(0.08^{+0.09}_{-0.07}\)&\(-0.19^{+0.02}_{-0.02}\)&0.00\\
\((0,\frac{3}{2})\) & \(\Xi_{cc}\pi\rightarrow\Xi_{cc}\pi\) & \(-0.101^{+0.003}_{-0.003}\) & \(13.28^{+1.04}_{-0.92}\)&\(-0.095^{+0.003}_{-0.004}\)&-0.08\\ 
\((-1,0)\) &   \(\Xi_{cc}\bar{K}\rightarrow\Xi_{cc}\bar{K}\)   &  \(-1.198^{+0.403}_{-1.307}\) &  \(0.02^{+0.12}_{-0.13}\) &\(-0.49^{+0.10}_{-0.19}\)&0.78\\ 
  & \(\Omega_{cc}\eta\rightarrow\Omega_{cc}\eta\)  &  \(-0.232^{+0.106}_{-0.128}\) \(+ i0.083^{+0.021}_{-0.021}\) & \(-0.25^{+0.16}_{-0.18}-i0.35^{+0.13}_{-0.17}\) &\(-0.26^{+0.03}_{-0.03}+i0.02^{+0.02}_{-0.01}\)&0.37 \\ 
  \((-1,1)\) & \(\Omega_{cc}\pi\rightarrow\Omega_{cc}\pi\) & \(0.014^{+0.010}_{-0.006}\) &  \(-27.72^{+16.48}_{-57.03}\)&\(0.03^{+0.01}_{-0.01}\)&0.04\\
  & \(\Xi_{cc}\bar{K}\rightarrow\Xi_{cc}\bar{K}\) & \(0.061^{+0.026}_{-0.021}+ i 0.094^{+0.015}_{-0.012}\) & \(0.68^{+1.87}_{-1.53}-i3.77^{+2.18}_{-2.29}\)&\(-0.22^{+0.14}_{-0.14}+i0.45^{+0.00}_{-0.09}\)&0.31\\
  \((0,\frac{1}{2})\)& \(\Xi_{cc}\pi\rightarrow\Xi_{cc}\pi\)&  \( 0.457^{+0.064}_{-0.047}\)&\(-8.05^{+0.45}_{-0.45}\)&\(0.55^{+0.16}_{-0.10}\)&0.29\\
   &  \(\Xi_{cc}\eta\rightarrow\Xi_{cc}\eta\)&\(0.106^{+0.050}_{-0.038}+ i0.075^{+0.051}_{-0.038}\) & \(-2.04^{+1.50}_{-2.82}-i2.51^{+2.06}_{-2.32}\)&\(-0.72^{+0.21}_{-0.17}+i0.30^{+1.10}_{-0.18}\)&0.32\\
   & \(\Omega_{cc}K\rightarrow\Omega_{cc}K\) &  \(-0.104^{+0.131}_{-0.098}+ i0.288^{+0.089}_{-0.084}\)&  \(-1.39^{+0.85}_{-0.69}-i0.82^{+0.90}_{-0.90}\)&\(-0.55^{+0.11}_{-0.16}+i0.13^{+0.19}_{-0.07}\)&0.56 \\
\hline  
\end{tabular}
\caption{Scattering lengths (third column) and effective ranges (fourth column) (in units of fm) calculated by taking the NLO UChPT expressions in Eq.~\eqref{eq.kcotduchpt} for both single and coupled channels at physical quark masses.} \label{tab.phya0r0} 
\end{table}

\subsection{Resonance poles in the complex energy plane}

The resonance poles can model-independently characterize the physical states in various reactions. In order to  obtain the resonance pole information, one needs to analytically extrapolate the scattering amplitudes to the complex energy plane on the unphysical Riemann sheets (RS's), where resonance poles reside. The continuation of the scattering amplitudes to complex energy plane can be achieved by extrapolating the $G(s)$ function to the second RS
\begin{equation}
\begin{aligned}
    G^{\Lambda}_{\rm II}(s)=G^{\Lambda}(s)-2i\rho(s)\,,
\end{aligned}
\end{equation}
where $\rho(s)$ is the kinematic factor defined previously and $G_{\rm II}^{\Lambda}(s)$ denotes the loop function on the second RS with ${\rm Im}G^{\Lambda}_{\rm II}(s)=-\rho(s)$ for $s>s_{\rm thr}$.

The scattering amplitudes with $n$ coupled channels will contain $2^n$ RS's. The first/physical RS can be denoted by $(+,+,+,\cdots,+)$, where each ``+'' symbol represents the sign of the imaginary part of the $G^{\Lambda}(s)$ function at the corresponding threshold. In this convention, the second, third and fourth RS's are given by $(-,+,+,\cdots,+)$, $(-,-,+,\cdots,+)$ and $(+,-,+,\cdots,+)$, respectively. Other RS's can be accessed via similar rules. 
The pole positions and their residues $(\gamma_i)$ on the complex energy plane for different scattering processes at physical quark masses are listed in Table~\ref{tab.pole}. 

\begin{table} [htbp]
\centering
\renewcommand{\arraystretch}{1.5}
\footnotesize
\begin{tabular}{c c  c c c} 
\hline 
\((S,I)\) & RS & Pole (MeV) & \(|\gamma_{i=1,2,3}|\text{(GeV)}\) & $X_{i=1,2,3}$\\
\hline  
\((1,0)\) & \( \text{II}\) & \(4055.9^{+19.9}_{-28.9}\) & (\(17.8^{+3.6}_{-2.0}\),--,--) & -- \\
\((-1,0)\) & \( \text{I}\) & \(4083.4^{+7.3}_{-7.5}\) & (\(17.8^{+0.6}_{-0.9}\), \(11.4^{+0.1}_{-0.3}\), --)& $(0.70^{+0.06}_{-0.05},\,0.06^{+0.00}_{-0.00})$\\
\((-1,1)\) & \( \text{II}\) & \(4066.9^{+6.1}_{-7.8}\) 
\(-i 354.7^{+31.2}_{-27.9}\) & (\(10.9^{+0.1}_{-0.1}\), \(26.4^{+2.7}_{-2.1}\),--)& $(0.22^{+0.00}_{-0.00},\,0.23^{+0.03}_{-0.02})$ \\
\((0,\frac{1}{2})\) & \( \text{II}\) & \(3828.9^{+0.8}_{-1.3}\) \(-i87.5^{+7.3}_{-6.8}\) & (\(15.9^{+0.1}_{-0.1}\), \(0.9^{+0.02}_{-0.01}\), \(6.8^{+0.4}_{-0.3}\)) & ($0.44^{+0.01}_{-0.01}$,\,$\approx 0$,\,$\approx 0$)\\
 & \( \text{IV}\) & $4146.4^{+16.6}_{-14.4}-i270.2^{+34.2}_{-31.3}$ & ($12.2^{+1.1}_{-0.5}$, $11.1^{+0.1}_{-0.0}$, $25.5^{+2.3}_{-1.7}$) &($0.10^{+0.01}_{-0.01}$,\,$0.25^{+0.01}_{-0.01}$,\,$0.25^{+0.02}_{-0.02}$)\\
& \( \text{II}\) & $4356.7^{+40.3}_{-39.1} -i152.5^{+1.1}_{-2.9}$ & ($6.8^{+0.8}_{-1.1}$, $15.9^{+0.5}_{-0.3}$, $20.6^{+0.4}_{-0.4}$) &($0.11^{+0.02}_{-0.03}$,\,$0.26^{+0.01}_{-0.00}$,\,$0.41^{+0.01}_{-0.01}$)\\
\hline  
\end{tabular}
\caption{ Pole positions, residues and the compositeness coefficients for the physical quark masses. }\label{tab.pole} 
\end{table} 

For the single-channel $\Xi_{cc}K$ scattering with $(S,I)=(1,0)$, the attraction is not strong enough to form a bound state and a virtual-state pole appears on the real axis of the second RS with the mass 4055.9 MeV, lying about 60 MeV below the $\Xi_{cc}K$ threshold at 4115.2~MeV. 
The upward phase shifts above the $\Xi_{cc}K$ threshold, as illustrated in Fig.~\ref{fig.scphaseshifts}, reflect the appearance of the virtual-state pole, consistent with the expectations from the studies in Refs.~\cite{Zheng:2003rw,Zhou:2004ms,Yao:2020bxx} revealing that the virtual pole would lead to rising positive values of phase shifts.

For the $\Xi_{cc}\bar{K}$ and $\Omega_{cc}\eta$ coupled channels with $(S,I)=(-1,0)$, we find a bound-state pole on the first RS lying about $32$ MeV below the $\Xi_{cc}\bar{K}$ threshold, with similar couplings to $\Xi_{cc}\bar{K}$ and $\Omega_{cc}\eta$.  
For the coupled-channel scattering of $\Omega_{cc}\pi$ and $\Xi_{cc}\bar{K}$ with $(S,I)=(-1,1)$, a very broad resonance pole, with a mass around 4067 MeV and a width around 700 MeV, appears on the second RS.  This resonance couples more strongly to the $\Xi_{cc}\bar{K}$ channel than to the $\Omega_{cc}\pi$ channel.

\begin{figure}[H]
  \centering
  \begin{subfigure}[b]{0.32\textwidth}
    \centering
    \includegraphics[width=\textwidth]{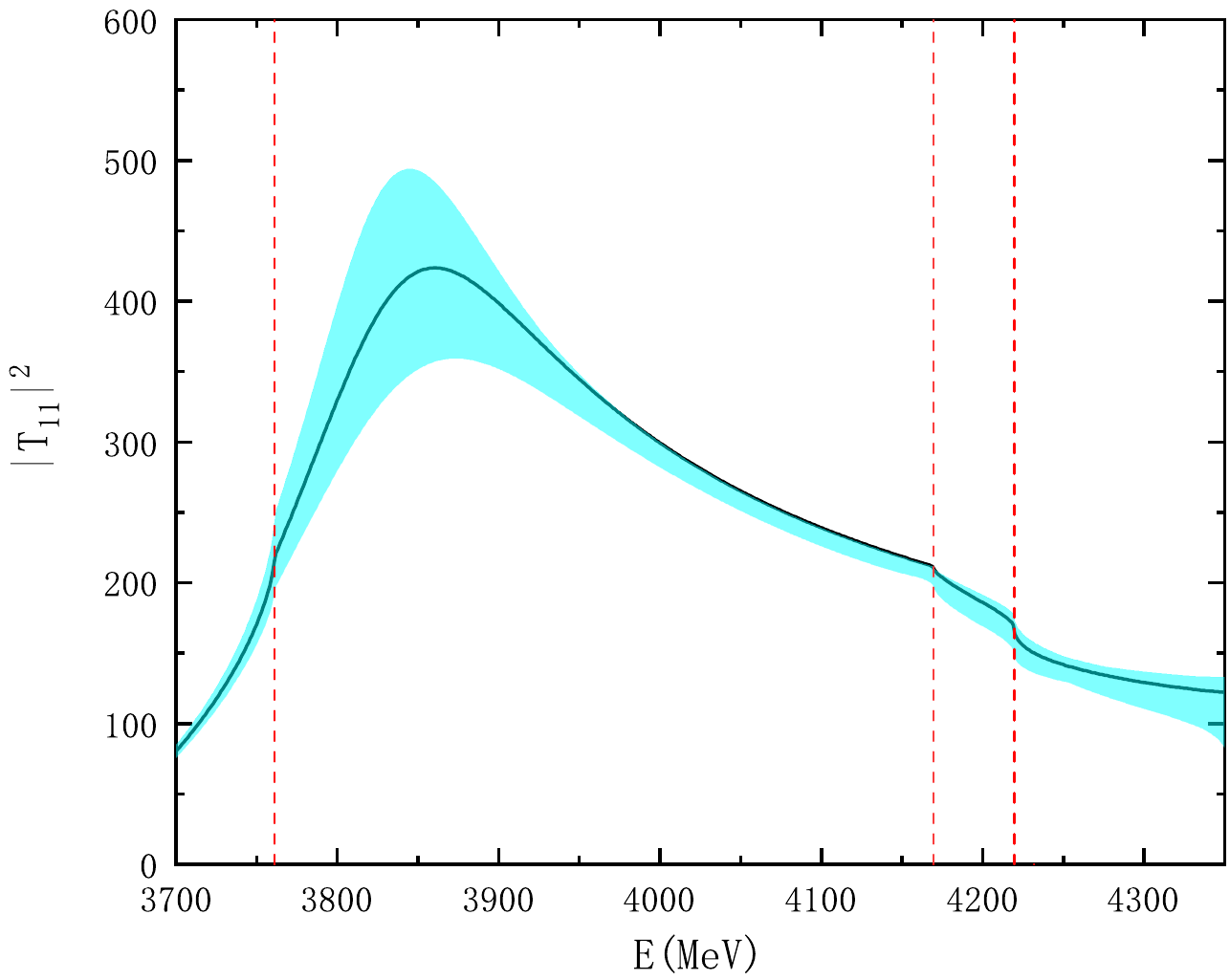}
  \end{subfigure}
  \begin{subfigure}[b]{0.33\textwidth}
    \centering
    \includegraphics[width=\textwidth]{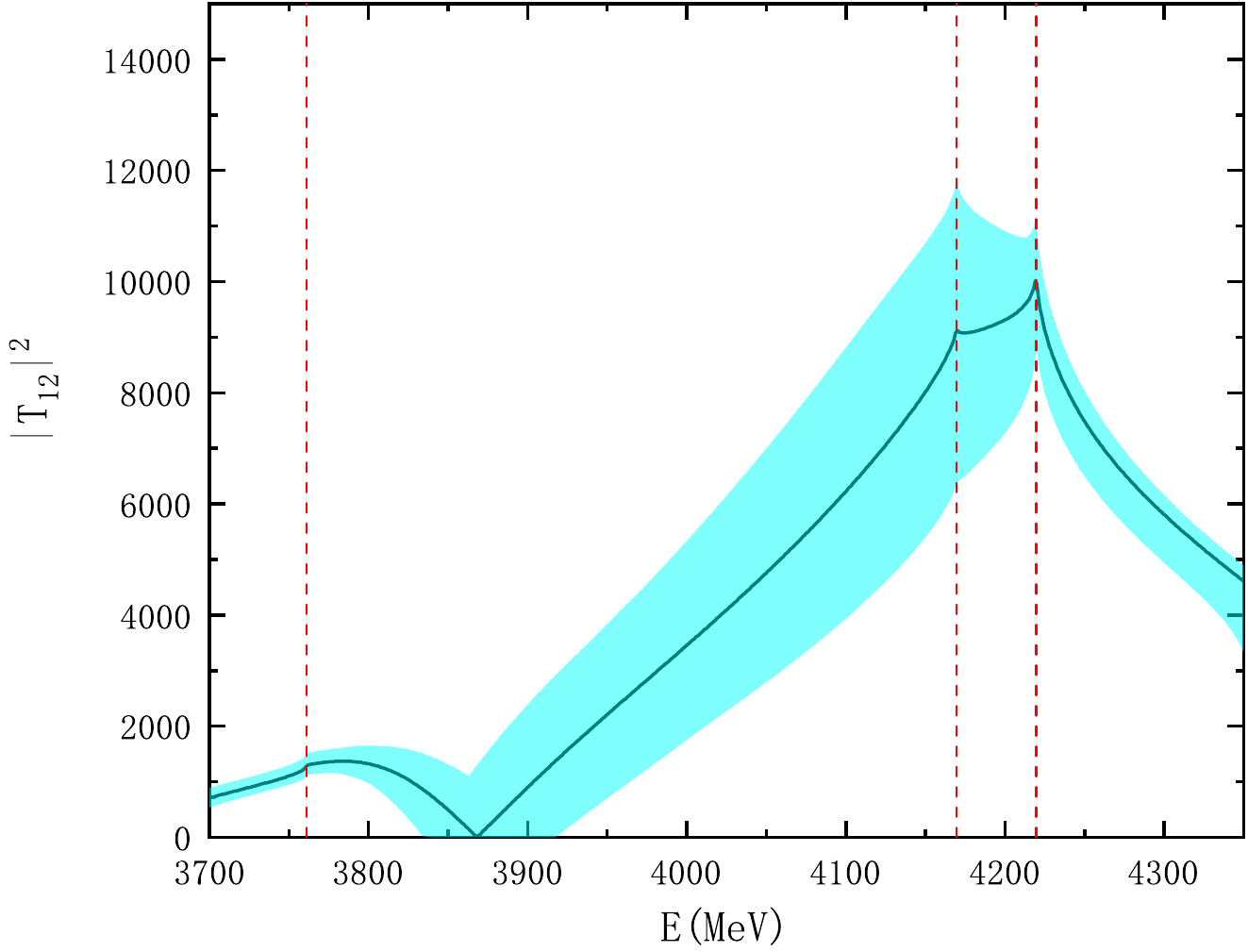}
  \end{subfigure}
  \begin{subfigure}[b]{0.33\textwidth}
    \centering
    \includegraphics[width=\textwidth]{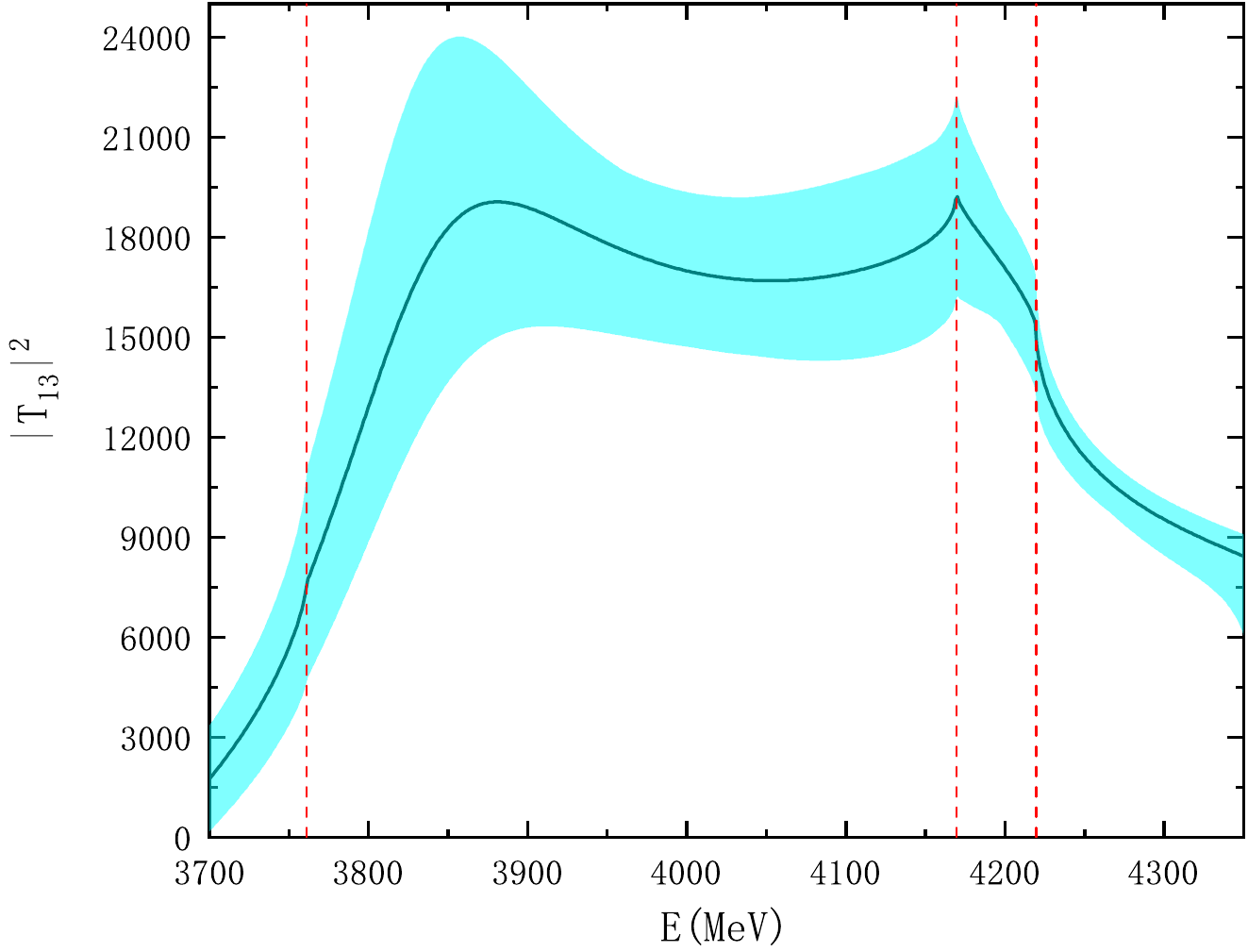}
  \end{subfigure}
  \caption{The magnitude squared of the scattering amplitudes with $(S,I)=(0,\frac{1}{2})$. $T_{11}$, $T_{12}$ and $T_{13}$ stand for the amplitudes of $\Xi_{cc}\pi\to\Xi_{cc}\pi$, $\Xi_{cc}\pi\to\Xi_{cc}\eta$ and $\Xi_{cc}\pi\to\Omega_{cc}K$, respectively. The vertical dashed lines from left to right in each panel correspond to the thresholds of $\Xi_{cc}\pi$, $\Xi_{cc}\eta$ and $\Omega_{cc}K$, in order. }\label{fig.T01d2}
\end{figure}

For the scattering processes with $(S,I)=(0,\frac{1}{2})$ that involves the channels of $\Xi_{cc}\pi$, $\Xi_{cc}\eta$ and $\Omega_{cc}K$, a resonance pole with a mass 3828.9 MeV and a half-width 87.5~MeV is located just above the $\Xi_{cc}\pi$ threshold at 3761.1~MeV on the second RS. According to the residues in Table~\ref{tab.pole}, this resonance couples most strongly to the $\Xi_{cc}\pi$ channel, while its coupling to $\Xi_{cc}\eta$ is very weak. This broad resonance pole is responsible for the prominent bump around 3.8~GeV in the $\Xi_{cc}\pi\to\Xi_{cc}\pi$ scattering amplitudes, as shown in Fig.~\ref{fig.T01d2}. The two resonance poles at around 4.15~GeV on the fourth RS and 4.36~GeV on the second RS are located distantly from the physical sheet. In fact there are also other distant resonance poles in the complex energy plane that we do not explicitly quote here. The effects of these distant poles are not quite visible in the physical scattering amplitudes, instead the threshold effects from the $\Omega_{cc}K$ and $\Xi_{cc}\eta$ channels are clearly observed, as illustrated in Fig.~\ref{fig.T01d2}.

The DCB resonance poles in the channels with $(S,I)=(-1,0)$ and $(0,1/2)$ are investigated by using a different theoretical framework in Ref.~\cite{Yan:2018zdt}, where preexisting DCB states with negative parity are explicitly included in the amplitudes. Tentative guesses are made for the bare masses and the bare couplings for the excited preexisting states. By taking the bare couplings as zero, the formalism of Ref.~\cite{Yan:2018zdt} reduces to the LO unitarized chiral approach as employed in Ref.~\cite{Guo:2017vcf}, which leads to qualitatively similar pole contents as obtained in the present work.

The compositeness relation, as first proposed by Weinberg~\cite{Weinberg:1962hj} to address the composite nature of the deuteron as a bound state of the proton and neutron, is possible to be generalized to describe the resonance state~\cite{Guo:2015daa}. We will follow the method in the latter reference to compute the compositeness coefficient $X_i$, which represents the probability of finding a molecular component in the $i$th two-body channel. The explicit expression of $X_i$ is given by~\cite{Guo:2015daa,Kang:2016ezb,Guo:2019kdc,Guo:2020pvt}:
\begin{equation}\label{eq.X}
\begin{aligned} 
X_i=|\gamma_i|^2\bigg\lvert\frac{dG^{\Lambda}_{i,({\rm II})}(s_R)}{ds}\bigg\rvert\,,
\end{aligned}
\end{equation}
where $\gamma_i$ is the residue of the resonance pole $s_R$ in the $i$th channel, and $dG^{\Lambda}_{i,({\rm II})}(s_R)/ds$ denotes the derivative of the function $G^{\Lambda}_{i,({\rm II})}(s)$ of the $i$th channel with respect to $s$ on the RS where the pole resides. It should be noted that Eq.~\eqref{eq.X} can be only applied to the poles on the complex energy plane lying above the corresponding channel thresholds , with the exception of the bound-state pole on the physical RS. The values of the compositeness coefficients for the various poles are shown in the last column of Table~\ref{tab.pole}. 

According the criteria of Ref.~\cite{Guo:2015daa}, the formula in Eq.~\eqref{eq.X} can not be applied to the virtual pole, which is the reason that we do not provide the $X$ value for the pole in the $(S,I)=(1,0)$ channel. 
For the resonance in the $(-1,1)$ channel, its mass of 4066.9 MeV is about 48 MeV below the $\Xi_{cc}\bar{K}$ threshold at 4115.2 MeV. Nevertheless, this resonance is very broad and its wave function still has an important distribution near threshold, allowing an approximate calculation of the compositeness coefficients. The compositeness coefficients for the two channels of $\Omega_{cc}\pi$ and $\Xi_{cc}\bar{K}$ are similar in magnitude, i.e., $X_1\simeq 0.22$, $X_2\simeq 0.23$, indicating that the resonance is a mixture of molecular and elementary types, with about $45\%$ of its molecular component originating from the $\Omega_{cc}\pi$ and $\Xi_{cc}\bar{K}$ two-body channels. 

For the resonance pole around 3.8~GeV appearing in the $(0,\frac{1}{2})$ coupled-channel case, its compositeness coefficient is dominated by the nearby $\Xi_{cc}\pi$ channel, with $X_{\Xi_{cc}\pi}\simeq 0.44$, while the distant channels of $\Xi_{cc}\eta$ and $\Omega_{cc}K$ give almost vanishing compositeness values. This implies that the resonance is a mixture of $\Xi_{cc}\pi$ molecular component and other heavier compact components beyond the three channels considered here. The resonance pole located around 4.1~GeV on the fourth RS and the one around 4.4~GeV on the second RS are not directly connected to the energy regions on the physical RS, implying that they will not lead to prominent enhancements in the physical regions, see Fig.~\ref{fig.T01d2} for the confirmation of this conclusion.

\section{Summary}\label{sec.sum}

In this work, we have focused on the next-to-leading order study of the ground doubly charmed baryons ($\Xi_{cc}^{++},\Xi_{cc}^{+},\Omega_{cc}^{+}$) and the light pseudoscalar mesons ($\pi,K,\eta$) scattering amplitudes within the chiral effective field theory. We perform the partial-wave projections of these scattering amplitudes. The unitarization procedure is implemented to properly include the possible non-perturbative interactions of the ground states of the doubly charmed baryons and light pseudoscalar mesons with definite strangeness and isospin quantum numbers. 

We fit the recent lattice data of $k\cot\delta$ in the four elastic scattering processes based on the CLQCD ensembles to fix the unknown next-to-leading order chiral low energy constants and the three-momentum cutoff introduced in the unitarization procedure. The predictions of the scattering lengths $a_0$ for various channels at the unphysical quark masses from our unitarized chiral amplitudes are found to be compatible with lattice determinations obtained from the effective-range-expansion method. While, the values of the effective ranges $r_0$ from the two approaches are not in good agreement in general. More precise lattice simulations are definitely helpful to resolve this issue. 

By performing the chiral extrapolation to physical quark masses, we have predicted phase shifts and inelasticities not only for the elastic scattering processes with $(S,I)=(-2,1/2)$, $(1,0)$,  $(1,1)$, $(0,3/2)$, but also for the coupled-channel scatterings with $(S,I)=(-1,0)$, $(-1,1)$ and $(0,1/2)$. A virtual-state pole is found to be located around 60~MeV below the $\Xi_{cc}K$ threshold in the elastic $\Xi_{cc}K$ scattering with $(S,I)=(1,0)$. A bound-state pole lying around 30~MeV below the $\Xi_{cc}\bar{K}$ threshold is found in the $\Xi_{cc}\bar{K}$ and $\Omega_{cc}\eta$ coupled-channel scattering with $(S,I)=(-1,0)$. For the coupled-channel scattering of $\Omega_{cc}\pi$ and $\Xi_{cc}\bar{K}$ with $(S,I)=(-1,1)$, we find a rather broad resonance pole lying around $(4.07-i0.36)$~GeV on the second Riemann sheet. For the three coupled channels of $\Xi_{cc}\pi$, $\Xi_{cc}\eta$ and $\Omega_{cc}K$ with $(S,I)=(0,1/2)$, we find a resonance pole at around $(3.83-i 0.09)$~GeV on the second RS, which leads to prominent enhancements in the scattering amplitudes on the physical real axis. Other broad resonance poles around $4.1$~GeV and $4.3$~GeV for the coupled-channel scattering with $(S,I)=(0,1/2)$ are found in the complex energy plane. But they do not give noticeable effects in the physical scattering amplitudes, since such poles at $4.1$~GeV and $4.3$~GeV are located on the Riemann sheets that are not directly connected to the physical sheet. 

It is expected that the results from our study can provide useful guidelines to future experimental measurements and lattice calculations for the excited doubly charmed baryons. 

\section*{Acknowledgements}

We thank Ze-Rui Liang, De-Liang Yao and Liuming Liu for useful discussions. This work is supported in part by Hebei Natural Science Foundation under Grants No.~A2026205016, by National Natural Science Foundation of China (NSFC) under Grants No.~12475078, and by the Science Foundation of Hebei Normal University with Contract No.~L2023B09. 

\bibliography{main}
\bibliographystyle{apsrev4-2}

\end{document}